\documentclass[a4paper,preprintnumbers,onecolumn,groupedaddress,showpacs,
nofootinbib,floatfix,superscriptaddress,aps,10pt]{revtex4-2}
\usepackage{graphicx}
\usepackage{epsfig}
\usepackage{bm}
\usepackage{amsfonts}
\usepackage[utf8]{inputenc}
\usepackage{amssymb}
\usepackage{subcaption}
\usepackage{amsmath}
\usepackage{dcolumn}
\usepackage{latexsym}
\usepackage{amsthm}
\usepackage{framed}
\usepackage{cancel}
\usepackage{mathtools}
\usepackage{tablefootnote}
\usepackage{multirow}
\usepackage[dvipsnames]{xcolor}
\usepackage[colorlinks]{hyperref}
\hypersetup{
    breaklinks=true,
    pdfstartview={FitH},    
    colorlinks=true,       
    linkcolor=blue,          
    citecolor=red,        
    filecolor=magenta,      
    urlcolor=magenta,           
    anchorcolor=green,      
    linktocpage=true
}

\begin{document}

\title{Phenomenological Rotating Extension of Black Holes with Primary Scalar Hair: Shadow Signatures in Beyond Horndeski Gravity}

\author{Kourosh Nozari}
\email[]{nozari7450@gmail.com, knozari@umz.ac.ir (Corresponding Author)}
\affiliation{Department of Theoretical Physics, Faculty of Science, University of Mazandaran,
P. O. Box 47416-95447, Babolsar, Iran}
\author{Milad Hajebrahimi}
\email[]{m.hajebrahimi@stu.umz.ac.ir}
\affiliation{Department of Theoretical Physics, Faculty of Science, University of Mazandaran,
P. O. Box 47416-95447, Babolsar, Iran}
\author{Sara Saghafi}
\email[]{s.saghafi@umz.ac.ir}
\affiliation{Department of Theoretical Physics, Faculty of Science, University of Mazandaran,
P. O. Box 47416-95447, Babolsar, Iran}
\affiliation{School of Physics, Damghan University, Damghan.
 3671645667, Iran}
\author{G. Mustafa}
\email[]{gmustafa@zjnu.edu.cn}
\affiliation{Department of Physics, Zhejiang Normal University, Jinhua 321004, China}
\affiliation{Research Center of Astrophysics and Cosmology, Khazar University, Baku, AZ1096, 41 Mehseti Street, Azerbaijan}
\author{Emmanuel N. Saridakis} \email{msaridak@noa.gr}
\affiliation{National Observatory of Athens, Lofos Nymfon, 11852 Athens, Greece}
\affiliation{Departamento de Matem\'{a}ticas, Universidad Cat\'{o}lica del Norte, Avda. Angamos 0610, Casilla 1280 Antofagasta, Chile}
\affiliation{CAS Key Laboratory for Researches in Galaxies and Cosmology, School of Astronomy and Space Science, University of Science and Technology of China, Hefei, Anhui 230026, China}

\begin{abstract}
The Event Horizon Telescope (EHT) image of M87* provides a direct test of 
strong-field gravity, measuring an angular shadow diameter $\theta_{d}=42\pm 
3~\mu\mathrm{as}$ and a circularity deviation $\Delta C\leq 0.1$. Such 
observations allow quantitative tests of the Kerr paradigm and of possible 
deviations from the no-hair theorem. In scalar-tensor extensions of gravity, 
black holes may possess primary scalar hair, introducing an additional 
independent parameter beyond mass and spin. In this work, we construct a 
rotating configuration inspired by black hole solutions with primary 
scalar hair in beyond Horndeski gravity and analyze their photon regions and 
shadow formation. We show that the scalar hair parameter $Q$ induces 
characteristic modifications of the shadow, and in particular negative $Q$ 
enlarges the shadow and reduces its oblateness, while positive $Q$ shrinks and 
enhances its distortion. Adopting M87* as a representative case
within this framework and imposing the EHT bounds on $\theta_{d}$ and $\Delta 
C$, we identify the viable $(a,Q)$ parameter space for the representative
fixed choice $\lambda/M=1/3$. We find that current 
observations do not exclude rotating black holes with primary scalar hair, 
although the allowed region is significantly restricted for $Q>0$. Finally, the 
scalar-hair-induced deviations are of order $\mathcal{O}(\mu\mathrm{as})$, 
placing them near the sensitivity threshold of present instruments and within 
reach of next-generation horizon-scale imaging.

\vspace{12 pt}

\textbf{Keywords}: Astrophysical black holes, Beyond Horndeski Theory, Event Horizon Telescope, Scalar hair

\end{abstract}

\maketitle


\section{Introduction}\label{sec:1}

In General Relativity (GR), the no-hair theorem states that stationary black hole solutions are uniquely characterized by a small number of conserved charges, namely mass, angular momentum, and electric charge \cite{Carter:1971zc,Israel:1967wq,Hawking:1971vc}. In realistic astrophysical environments, however, black holes are expected to carry negligible electric charge, since any accumulated charge is efficiently neutralized by surrounding plasma and radiation fields \cite{Gibbons:1975jb,Zajacek:2018ycb}. Consequently, rotating astrophysical black holes are well described by the Kerr metric, which represents the unique asymptotically flat, stationary, vacuum solution of GR under these assumptions.

Despite its remarkable empirical success across a wide range of scales, GR is believed to be incomplete. From a theoretical perspective, higher-derivative extensions of GR arising in renormalization schemes generically introduce ghost instabilities \cite{Stelle:1976gc}. On the observational side, various cosmological tensions \cite{CosmoVerseNetwork:2025alb} and the evidence for dark matter and dark energy from cosmological surveys \cite{Zwicky:1933gu,SDSS:2014irn,AtacamaCosmologyTelescope:2013swu,BOSS:2012dmf}, indicate that either new matter components or modifications of gravitational dynamics may be required. These considerations have motivated extensive investigations of modified theories of gravity.

Gravitational modifications \cite{CANTATA:2021asi} arise by extending the Einstein-Hilbert lagrangian in various ways, resulting in $f(R)$ gravity \cite{Starobinsky:1980te}, $f(G)$ gravity \cite{Nojiri:2005jg}, Lovelock gravity \cite{Lovelock:1971yv}, etc. Nevertheless, one can be based on the torsional formulation and extend it to $f(T)$ gravity \cite{Cai:2015emx}, to $f(T,T_{G})$ gravity \cite{Kofinas:2014owa}, to $f(T,B)$ gravity \cite{Bahamonde:2015zma}, etc. Similarly, one can start from the equivalent formulation of gravity in terms of non-metricity, and construct $f(Q)$ gravity \cite{Heisenberg:2023lru}, $f(Q,C)$ gravity \cite{De:2023xua}, etc.  

Among alternative frameworks, scalar-tensor theories play an important role, since they represent the most direct and well-controlled extension of GR through the inclusion of an additional scalar degree of freedom. The most general scalar-tensor theory leading to second-order field equations is Horndeski gravity \cite{Horndeski:1974wa,DeFelice:2010nf,Deffayet:2013lga,Cisterna:2014nua,Kobayashi:2014ida,Bellini:2015xja,Babichev:2016rlq,Kase:2018aps,Banerjee:2018svi,
Kobayashi:2019hrl,Kovacs:2020ywu,Bahamonde:2021dqn,Lu:2020iav,Petronikolou:2021shp,Mandal:2023kpu,Santos:2023eqp,Rayimbaev:2023bjs,Santos:2024cvx,Chen:2025aom,
Hoshimov:2025hyt,Santos:2025fdp,Mironov:2025dzz,Totolou:2025dsb}. It was later realized, however, that the Horndeski construction does not exhaust all healthy scalar-tensor models. In particular, beyond Horndeski (or GLPV) theories extend the Horndeski framework while preserving the correct number of propagating degrees of freedom through hidden constraints, despite containing higher-order derivatives \cite{Gleyzes:2014dya,Zumalacarregui:2013pma,Deffayet:2015qwa}. These theories have attracted considerable interest due to their theoretical consistency and rich phenomenology across cosmological and astrophysical scales \cite{Babichev:2016jom,Crisostomi:2016tcp,Sakstein:2016ggl,Sakstein:2016oel,BenAchour:2016cay,Langlois:2015skt,BenAchour:2016fzp,Dima:2017pwp,Kolevatov:2017voe,
Kase:2018iwp,Zhu:2021whu}.

A distinctive feature of certain subclasses of beyond Horndeski gravity is the existence of black hole solutions endowed with genuine \emph{primary scalar hair}. Contrary to secondary hair, which is completely determined by the black hole mass and angular momentum, primary scalar hair corresponds to an independent integration constant associated with the scalar field configuration. Recently, Bakopoulos \textit{et al.} constructed explicit static, spherically symmetric black hole solutions with primary scalar hair within shift- and parity-symmetric beyond Horndeski theories \cite{Bakopoulos:2023fmv}. These solutions represent genuine departures from the Kerr geometry while remaining regular and asymptotically flat, thereby providing a well-defined framework for exploring strong-field signatures of scalar-tensor gravity. Related black hole configurations within Horndeski and beyond Horndeski gravity have been studied in various contexts, including black hole thermodynamics, gravitational lensing, and solar-system phenomenology \cite{Rinaldi:2012vy,Anabalon:2013oea,Minamitsuji:2013ura,Sotiriou:2014pfa,Maselli:2015yva,Mukherjee:2017fqz}.

As it is well known, black holes constitute natural laboratories for testing gravity in the strong-field regime. A major observational milestone in this direction has been achieved by the Event Horizon Telescope (EHT) collaboration, which obtained horizon-scale images of the supermassive black holes at the centers of M87 and the Milky Way \cite{EventHorizonTelescope:2019dse,EventHorizonTelescope:2019uob}. Using Very Long Baseline Interferometry (VLBI), the EHT resolved a central brightness depression interpreted as the black hole shadow, produced by strong gravitational lensing of photons emitted by the surrounding plasma. The shadow of M87* is broadly consistent with the Kerr prediction, exhibiting an angular diameter of $42\pm 3~\mu\mathrm{as}$, a deviation from circularity $\Delta C\leq 10\%$, and an axis ratio $\lesssim 4/3$ \cite{EventHorizonTelescope:2019dse,EventHorizonTelescope:2019uob}. Nevertheless, the present uncertainties, including limitations in array coverage and in modeling of accretion and emission processes, leave room for moderate deviations from Kerr geometry \cite{Gralla:2020pra}.

Within GR, the shadow of a rotating black hole is determined solely by its mass, spin, and the observer's inclination angle. However, in modified gravity theories   additional degrees of freedom can modify null geodesics and consequently alter the size, shape, or displacement of the shadow, even in stationary and asymptotically flat spacetimes \cite{Johannsen:2013szh,Konoplya:2016hmd,Cardoso:2019rvt}. Black hole shadows therefore provide a direct probe of the spacetime geometry and, by extension, of the underlying gravitational theory \cite{Falcke:1999pj,Johannsen:2015mdd,Bambi:2015kza,Goddi:2016qax,Khodadi:2020gns,EventHorizonTelescope:2020qrl,Jusufi:2021fek}.

Motivated by these considerations, in the present work we construct a
stationary and axisymmetric rotating extension of the exact static black-hole
solution with primary scalar hair found in Ref.~\cite{Bakopoulos:2023fmv},
using the revised Newman--Janis procedure
\cite{Azreg-Ainou:2014pra,Drake2000}. We emphasize from the outset that,
in a generic modified-gravity theory, the Newman--Janis algorithm is not a
solution-generating transformation of the underlying field equations.
Consequently, although the static seed metric and its scalar profile constitute
an exact solution of the beyond-Horndeski theory considered here, the rotating
metric obtained from the revised Newman--Janis construction should not be
regarded, on the basis of the algorithm alone, as an exact rotating solution
of the same theory.

In Appendix~\ref{apapc2}, we investigate this issue explicitly. We show that
the original static scalar ansatz $\Phi=qt+\Psi(r)$ is incompatible with the
rotating geometry and derive the equations that a more general stationary and
axisymmetric scalar configuration
$\widetilde{\Phi}=qt+\widetilde{\Psi}(r,\theta)$ would have to satisfy.
However, we do not provide an existence proof or a numerical solution of the
resulting coupled system. We therefore adopt the rotating metric in the
remainder of this work as a phenomenological Kerr-like extension continuously
connected to the exact static beyond-Horndeski solution and to the Kerr limit.
Accordingly, the shadow bounds obtained below constrain this rotating metric
parametrization and should not be interpreted as direct constraints on an
established exact rotating solution of the underlying beyond-Horndeski theory.

In particular, adopting M87* as a representative case within this
phenomenological framework, we impose the observational bounds
$\Delta C\leq 0.1$ and
$39~\mu\mathrm{as}\leq\theta_{d}\leq 45~\mu\mathrm{as}$ on the parameter
space spanned by the spin $a$ and the deformation parameter $Q$, where $Q$
originates from the primary scalar hair of the exact static seed solution.
While we do not perform a full parameter-inference analysis, our goal is to
identify the qualitative and quantitative effects of this hair-induced
deformation on the shadow morphology and to examine its potential degeneracy
with rotation. The resulting bounds should therefore be understood as
constraints within the phenomenological rotating metric considered here,
rather than as direct constraints on an exact rotating solution of the
underlying beyond-Horndeski theory. This approach is in line with the general
strategy adopted in studies of black-hole shadows based on phenomenological
or parametrized deviations from Kerr
\cite{Amarilla2012,Cunha2018,Kumar:2020hgm}.

The manuscript is organized as follows. In Sec.~\ref{sec:2}, we present 
the rotating configuration with primary scalar hair in beyond Horndeski 
gravity and analyze its horizon structure. In Sec.~\ref{sec:3}, we investigate 
photon motion and derive the associated photon regions and shadow observables. 
Section~\ref{sec:4} discusses the phenomenological implications of the model 
and analyzes the sensitivity of shadow observables to the underlying 
parameters, including constraints from EHT observations of M87*. Finally, in 
Sec.~\ref{sec:5}, we summarize our results and outline future observational 
prospects.

\section{Black Holes with Primary Scalar Hair}\label{sec:2}

In this section, we present the gravitational configuration that forms the
basis of our analysis. We begin by reviewing the exact spherically symmetric
black-hole solution with primary scalar hair constructed in
Ref.~\cite{Bakopoulos:2023fmv} within a subclass of beyond Horndeski theories \cite{Gleyzes:2014dya,Crisostomi:2016tcp}.
We then apply the revised Newman--Janis procedure \cite{Azreg-Ainou:2014pra,Azreg-Ainou:2014aqa} to construct a stationary
and axisymmetric phenomenological rotating extension of this static geometry.
As discussed below and in Appendix~\ref{apapc2}, this procedure does not by
itself establish the resulting metric as an exact rotating solution of the
underlying beyond-Horndeski field equations. The resulting metric nevertheless
provides a well-defined geometrical framework for investigating the effects of
the hair-induced deformation on photon motion and shadow observables.

\subsection{Spherically symmetric black hole in beyond Horndeski framework}

The action of beyond Horndeski 
theories~\cite{Gleyzes:2014dya,Crisostomi:2016tcp} with a scalar field $\Phi$ 
possessing the additional shift symmetry $\Phi\to\Phi+\mathrm{const.}$ and the 
parity symmetry $\Phi\to-\Phi$ can be read as follows
\begin{eqnarray}\label{action}
&&\mathcal{I}\left[\tilde{g}_{\mu\nu},\Phi\right]=\frac{1}{2\kappa}\int\mathrm{d
}^4x\sqrt{-\tilde{g}}\, 
\Bigl\{G_{2}\left(W\right)+G_{4}\left(W\right)\tilde{R}+\partial_{W}G_{4}
\left(W\right)\left[\left(\Box\Phi\right)^{2}-\left(\nabla_{\mu}
\partial_{\nu}\Phi\right)\left(\nabla^{\mu}\partial^{\nu}\Phi\right)\right] 
\nonumber\\ 
&&
+F_{4}\left(W\right)\epsilon^{\mu\nu\rho\sigma}\epsilon^{\alpha\beta\gamma}_{
\hspace{0.5 
cm}\sigma}\,\left(\partial_{\mu}\Phi\right)\left(\partial_{\alpha}
\Phi\right)\left(\nabla_\nu\partial_\beta\Phi\right)
\left(\nabla_\rho\partial_\gamma\Phi\right)\Bigl\}\,,
\end{eqnarray}
where $\kappa=8\pi G_{N}/c^{4}$ is the coupling constant, $G_{N}$ is the 
Newtonian constant, $c$ is the speed of light (from now on, we implement the 
natural units $G_{N}=\hbar=c=1$), $\Box=\partial^{\mu}\partial_{\mu}$ is the 
d'Alembertian operator, $\tilde{g}$ is determinant of the background spacetime 
metric $\tilde{g}_{\mu\nu}$ with the Ricci scalar $\tilde{R}$, and $G_{2}(W)$, 
$G_{4}(W)$, and $F_{4}(W)$ are three parametrization functions in terms of the 
scalar field kinetic expression 
$W=-\frac{1}{2}\partial_{\mu}\Phi\,\partial^{\mu}\Phi$. Moreover, the additional 
shift symmetry in the theories causes a time-dependency for the scalar field to 
the form of
\begin{equation}\label{scfi}
\Phi=qt+\Psi(r)\,,
\end{equation}
where $\Psi(r)$ can be found in Ref.~\cite{Bakopoulos:2023fmv}. In expression 
\eqref{scfi}, the quantity $q$ plays the role of the primary scalar hair with 
the dimension of $[\mathrm{length}]^{-1}$ to ensure the dimensionless behavior 
of the scalar field $\Phi$.

In order to construct an asymptotically flat black hole endowed with primary 
scalar hair within shift-symmetric beyond Horndeski gravity, we follow the 
approach of Ref. \cite{Bakopoulos:2023fmv} and we specify the free functions of 
the theory accordingly. In particular, we choose
\begin{equation}\label{the1}
G_{2}=-\left(\frac{8\eta}{3\lambda^{2}}\right)W^{2}\,,\qquad 
G_{4}=1-\left(\frac{4\eta}{3}\right)W^{2}\,,\qquad F_{4}=\eta\,,
\end{equation}
where $\lambda$ and $\eta$ are coupling constants with dimensions of 
$[\mathrm{length}]$ and $[\mathrm{length}]^{4}$, respectively. Without loss of 
generality, we take $\lambda>0$, since the theory is invariant under the 
transformation $\lambda\rightarrow-\lambda$, as can be directly verified from 
\eqref{the1}. On the other hand, the coupling $\eta$ may take either positive or 
negative values, leading to qualitatively different behaviors of the resulting 
solutions.

Solving the field equations corresponding to the action \eqref{action} with the 
above choice of functions, one obtains a static and spherically symmetric black 
hole solution of the form
\begin{equation}\label{ssle}
\mathrm{d}s^{2}=-f(r)\,\mathrm{d}t^{2}+\frac{\mathrm{d}r^{2}}{f(r)}+r^{2}\mathrm
{d}\Omega^{2}\,,
\end{equation}
where 
$\mathrm{d}\Omega^{2}=\mathrm{d}\theta^{2}+\sin^{2}\theta\,\mathrm{d}\varphi^{2}
$ is the line element of unit two-sphere. The metric function $f(r)$ is given by
\begin{equation}\label{fbh1}
f(r)=1-\frac{2M}{r}+Q\left[\frac{1}{1+(r/\lambda)^{2}}+\frac{\frac{\pi}{2}
-\arctan(r/\lambda)}{(r/\lambda)}\right]\,,
\end{equation}
where we have defined the effective deformation parameter
\begin{equation}
Q\equiv \eta q^{4}.
\end{equation}
Here, $M$ corresponds to the Arnowitt--Deser--Misner (ADM) mass of
the black hole, whereas $q$ is an independent integration constant
associated with the scalar field and characterizes the primary scalar
hair. Since $q^{4}\geq0$, the sign of $Q$ is determined entirely by
the sign of the coupling $\eta$,
\begin{equation}
\operatorname{sgn}(Q)=\operatorname{sgn}(\eta),
\qquad q\neq0.
\end{equation}
Consequently, the branches $Q<0$ and $Q>0$ correspond respectively
to $\eta<0$ and $\eta>0$, independently of the sign of the scalar
charge $q$. At fixed $\eta$, the magnitude of the deformation is
related to the magnitude of the scalar charge through
\begin{equation}
|Q|=|\eta|\,|q|^{4}.
\end{equation}
In the following numerical analysis, $Q$ is therefore treated as the
effective metric deformation parameter, while statements concerning
variations of $q$ are made only when the coupling $\eta$ is explicitly
held fixed.

The corresponding scalar field configuration reads \cite{Bakopoulos:2023fmv}
\begin{equation}\label{Phibh1}
\Phi(t,r)=qt+\Psi(r)\,,\qquad 
\left[\Psi'(r)\right]^{2}=\frac{q^{2}}{f^{2}(r)}\left(1-\frac{f(r)}{
1+(r/\lambda)^{2}}\right)\,,
\end{equation}
where a prime denotes differentiation with respect to $r$. The associated 
kinetic term of the scalar field is then found to be
\begin{equation}\label{Xbh1}
W=\frac{q^{2}/2}{1+(r/\lambda)^{2}}\,.
\end{equation}

As evident from the above expressions, the solution possesses two independent 
integration constants, namely the mass parameter $M$ and the scalar charge $q$, 
the latter being unrelated to $M$ and thus characterizing primary scalar hair. 
In the limit $q\rightarrow 0$, or equivalently $Q\rightarrow 0$, the scalar 
field becomes trivial and the metric function \eqref{fbh1} reduces smoothly to 
that of the Schwarzschild black hole, as expected.

The horizons of the black hole are determined by the real positive roots of 
$f(r)=0$. Due to the functional form of $f(r)$, closed analytic expressions for 
the horizon radii cannot be obtained. Depending on the values of $q$, $\eta$, 
and $\lambda$, the spacetime may exhibit a single event horizon, multiple 
horizons, or a naked singularity, as discussed in detail in 
Ref.~\cite{Bakopoulos:2023fmv}. In the present work, we restrict ourselves to 
the physically relevant parameter region for which the solution describes a 
black hole with a single event horizon $r_{\mathrm{eh}}$.

\begin{figure}[htb]
\centering
	\begin{tabular}{cc}
		\includegraphics[width=0.4\textwidth]{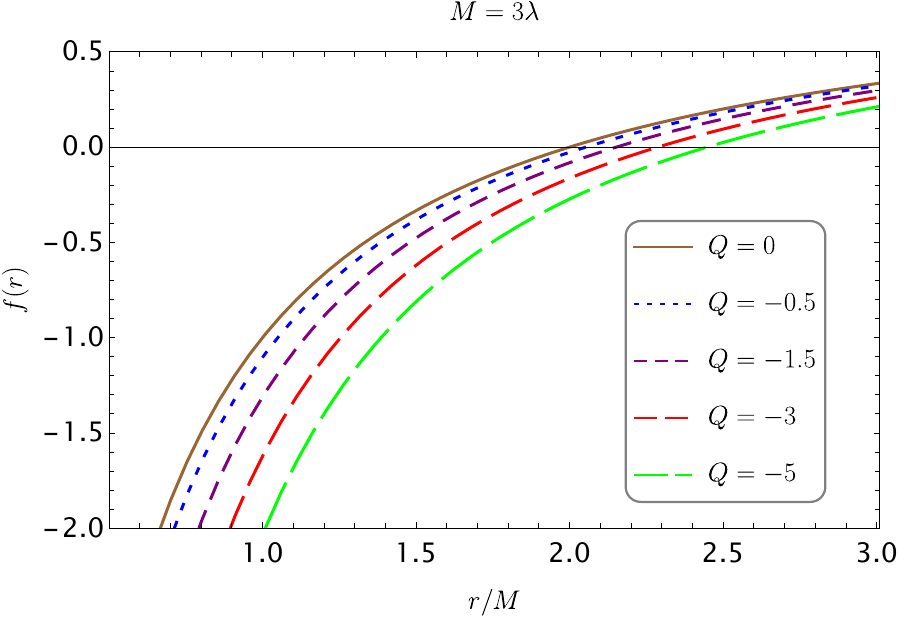}\, 
		\includegraphics[width=0.4\textwidth]{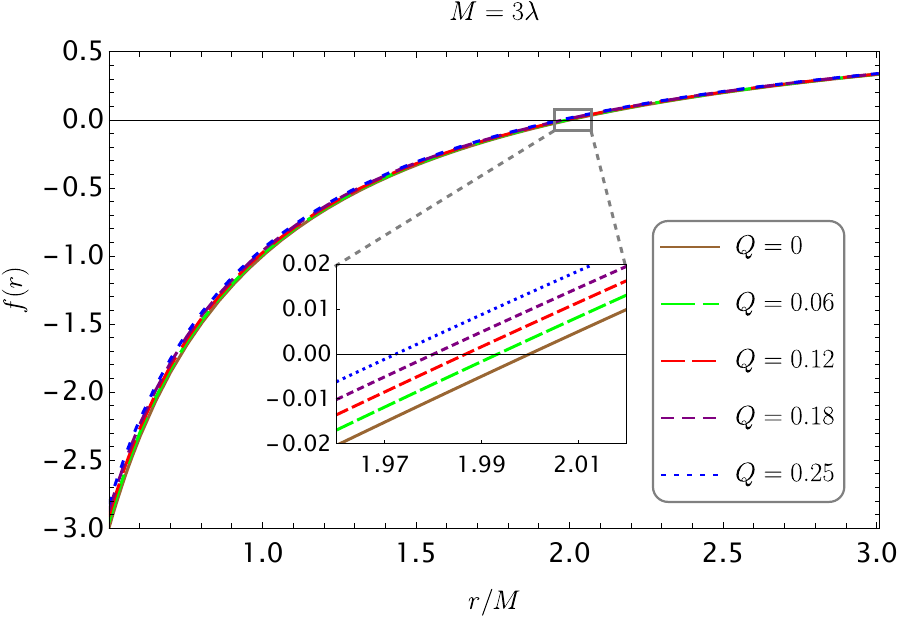}\,  \\
		 \includegraphics[width=0.4\textwidth]{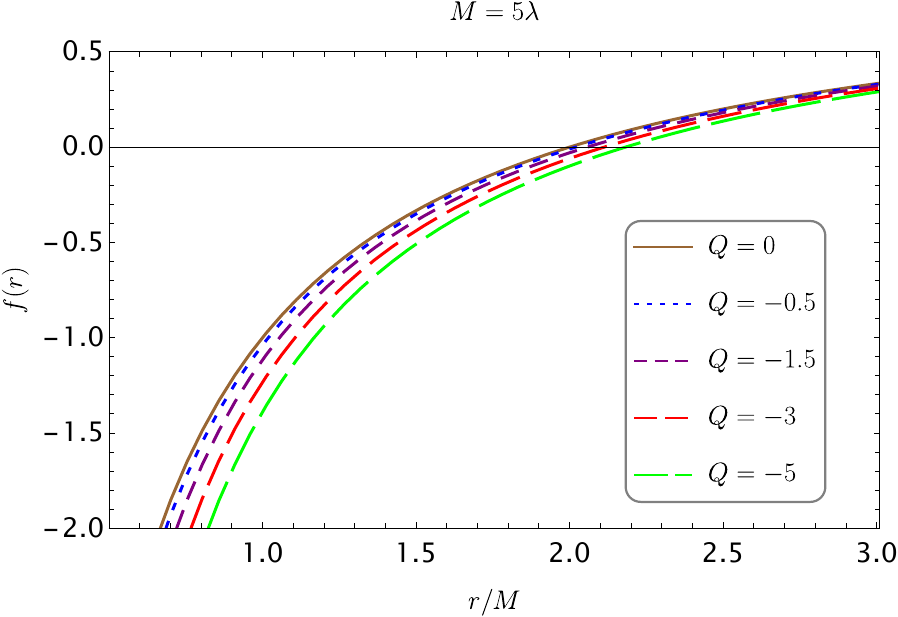}\,  
		\includegraphics[width=0.4\textwidth]{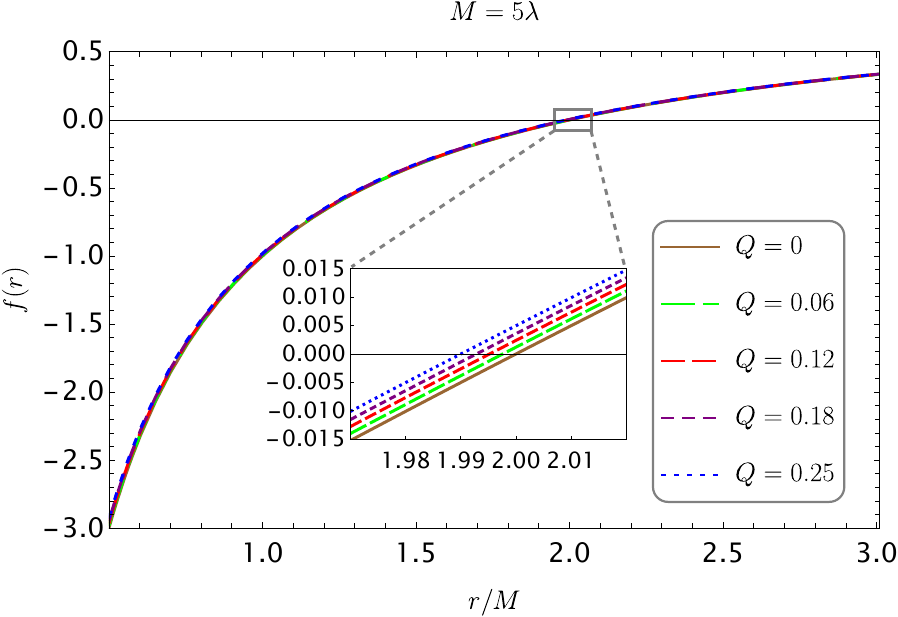}
	\end{tabular}
\caption{\label{Fig1}\small{\emph{The radial behavior of the metric function 
$f(r)$. The upper panels correspond to $M=3\lambda$ and the lower panels to 
$M=5\lambda$, while the left panels correspond to $Q<0$ and the right panels to 
$Q>0$.}}}
\end{figure}

In Fig.~\ref{Fig1} we display the radial behavior of the metric function $f(r)$ 
for representative values of $M$ and $Q$. As we observe, for $Q<0$ 
(corresponding to $\eta<0$), increasing the magnitude of $|Q|$ leads to an 
outward shift of the event horizon, whereas for $Q>0$ (i.e., $\eta>0$) the 
horizon radius decreases as $Q$ increases. Furthermore, for $M=5\lambda$ we 
observe a more compact configuration due to the larger mass.

For completeness, in Table~\ref{Table1}, we summarize the numerical values of 
the event horizon radius for representative choices of the parameters. As 
expected, for fixed $Q$ the horizon radius increases with the mass parameter 
$M$, while for fixed $M$ the presence of scalar hair modifies the horizon 
location in a manner controlled by the sign and magnitude of $Q$.
\begin{table}[htb]
\centering
\caption{\label{Table1}\small{\emph{The values of the event horizon radius 
$r_{\mathrm{eh}}$ of the asymptotically flat black hole with primary scalar 
hair, for representative choices of the parameters. The case $Q=0$ corresponds 
to the Schwarzschild limit with $r_{\mathrm{Sch}}=2$.}}}
\begin{tabular}{c|c|cccc|cccc|c}
\multirow{2}{*}{horizons} & Sch   & \multicolumn{4}{c|}{$Q<0$}                  
                                                           & 
\multicolumn{4}{c|}{$Q>0$}                                                      
                         & \multirow{2}{*}{$\lambda$ values} \\ \cline{2-10}
                          & $Q=0$ & \multicolumn{1}{c|}{$Q=-0.5$} & 
\multicolumn{1}{c|}{$Q=-1.5$} & \multicolumn{1}{c|}{$Q=-3$}  & $Q=-5$  & 
\multicolumn{1}{c|}{$Q=0.06$} & \multicolumn{1}{c|}{$Q=0.12$} & 
\multicolumn{1}{c|}{$Q=0.18$} & $Q=0.25$ &                                   \\ 
\hline\hline
$r_{\mathrm{eh}}$         & $2$   & \multicolumn{1}{c|}{$2.053$}  & 
\multicolumn{1}{c|}{$2.152$}  & \multicolumn{1}{c|}{$2.287$} & $2.448$ & 
\multicolumn{1}{c|}{$1.993$}  & \multicolumn{1}{c|}{$1.986$}  & 
\multicolumn{1}{c|}{$1.980$}  & $1.972$  & $M=3\lambda$                      \\ 
\hline
$r_{\mathrm{eh}}$         & $2$   & \multicolumn{1}{c|}{$2.019$}  & 
\multicolumn{1}{c|}{$2.057$}  & \multicolumn{1}{c|}{$2.112$} & $2.182$ & 
\multicolumn{1}{c|}{$1.997$}  & \multicolumn{1}{c|}{$1.995$}  & 
\multicolumn{1}{c|}{$1.992$}  & $1.990$  & $M=5\lambda$                     
\end{tabular}
\end{table}

\subsection{Rotating black hole configuration with primary scalar hair}

The well-known method for converting non-rotating black hole solutions into
rotating counterparts is the Newman--Janis algorithm (NJA)~\cite{Newman:1965tw},
which was originally formulated within General Relativity in order to generate
the Kerr black hole from the Schwarzschild seed metric. However, the ordinary
NJA generally fails to transform the resulting rotating metric from
Eddington--Finkelstein coordinates (EFCs) to Boyer--Lindquist coordinates
(BLCs) through real and globally integrable coordinate transformations, mainly
due to the complexification procedure imposed on the radial coordinate
$r$~\cite{Hansen:2013owa}. A revised version of the NJA, designed to overcome
this difficulty through a non-complexification prescription, was proposed in
Refs.~\cite{Azreg-Ainou:2014pra,Azreg-Ainou:2014aqa}. Both the ordinary and
revised NJA have subsequently been employed in many black hole solutions within
modified gravity theories~\cite{Johannsen:2011dh,Jusufi:2019caq,Bambi:2013ufa,
Moffat:2014aja}.

Motivated by this approach, we apply the revised NJA to the static line element
\eqref{ssle} in order to construct a rotating configuration endowed with
primary scalar hair in beyond Horndeski gravity. Consequently, the resulting
stationary and axisymmetric metric in BLCs can be written as (see Appendix
\ref{apap1} for technical details)
\begin{equation}\label{rle}
\mathrm{d}s^{2}=
-\left(1-\frac{2\rho(r)}{\Sigma}\right)\mathrm{d}t^{2}
+\frac{\Sigma}{\Delta}\,\mathrm{d}r^{2}
-\frac{4a\rho(r)\sin^{2}\theta}{\Sigma}\,\mathrm{d}t\,\mathrm{d}\varphi
+\Sigma\,\mathrm{d}\theta^{2}
+\sin^{2}\theta\left(a^{2}+r^{2}
+a^{2}\sin^{2}\theta\frac{2\rho(r)}{\Sigma}\right)\mathrm{d}\varphi^{2},
\end{equation}
where $a$ denotes the spin parameter, while
\begin{equation}\label{defle}
\Sigma=r^{2}+a^{2}\cos^{2}\theta,\qquad
\Delta=a^{2}+r^{2}f(r),\qquad
2\rho(r)=a^{2}+r^{2}-\Delta.
\end{equation}

The above rotating geometry is generated from the same static seed solution of
the beyond Horndeski action \eqref{action}. Nevertheless, once rotation is
introduced, the scalar sector supporting the static solution must also be
generalized. In particular, if one keeps the original scalar-field ansatz
\eqref{scfi}, the associated kinetic term
$W=-\frac{1}{2}\partial_{\mu}\Phi\partial^{\mu}\Phi$ is no longer solely a
function of $r$, but acquires angular dependence through the rotating metric.
As a consequence, certain off-diagonal components of the modified field
equations, especially $\mathcal{E}_{r\theta}$ and
$\mathcal{E}_{t\theta}$, no longer vanish identically, and the static scalar
ansatz ceases to provide an exact rotating solution.

Therefore, in order to consistently support the rotating geometry while keeping
the underlying beyond Horndeski theory unchanged, the scalar field must be
promoted to a stationary and axisymmetric configuration of the form
\begin{equation}\label{newsca-main}
\tilde{\Phi}(t,r,\theta)=qt+\tilde{\Psi}(r,\theta),
\end{equation}
where the time-linear term preserves the same conserved scalar charge $q$,
while the function $\tilde{\Psi}(r,\theta)$ dynamically adjusts to the
rotational deformation of spacetime. In this way, the scalar field remains
compatible with stationarity and axial symmetry, while its level surfaces are
deformed from spherical to axisymmetric ones.

The corresponding kinetic term becomes
\begin{equation}\label{wtilde-main}
\tilde{W}(r,\theta)=
-\frac{1}{2}g^{\mu\nu}
\partial_{\mu}\tilde{\Phi}\,
\partial_{\nu}\tilde{\Phi},
\end{equation}
Using the inverse metric components
\begin{equation}
g^{tt}
=
-\frac{(r^{2}+a^{2})^{2}-a^{2}\Delta\sin^{2}\theta}
{\Sigma\Delta},
\qquad
g^{rr}=\frac{\Delta}{\Sigma},
\qquad
g^{\theta\theta}=\frac{1}{\Sigma},
\end{equation}
and noting that $\partial_{\phi}\widetilde{\Phi}=0$, the definition
\ref{wtilde-main} gives
\begin{equation}\label{PDE-main}
\Delta(\partial_{r}\widetilde{\Psi})^{2}
+(\partial_{\theta}\widetilde{\Psi})^{2}
-\frac{(r^{2}+a^{2})^{2}
-a^{2}\Delta\sin^{2}\theta}{\Delta}\,q^{2}
+2\widetilde{W}\Sigma=0.
\end{equation}

As a check, in the static limit $a\to0$, for which
$\Sigma\to r^{2}$ and $\Delta\to r^{2}f(r)$, Eq.~\eqref{PDE-main}
reduces to
\begin{equation}
f(r)\,[\Psi'(r)]^{2}
-\frac{q^{2}}{f(r)}
+2W(r)=0,
\end{equation}
which is consistent with the kinetic term of the exact static scalar
configuration given in Eqs.~\eqref{scfi}--\eqref{Xbh1}.

Moreover, the nontrivial components of the gravitational field equations reduce
to a coupled system of two-dimensional partial differential equations. In
particular, the $(r\theta)$-component gives
\begin{equation}\label{ertheta-main}
\mathcal{E}_{r\theta}
=
\eta\sin^{2}\theta
\left[
\Delta(\partial_{r}\partial_{\theta}\tilde{\Psi})
-\Sigma(\partial_{r}\tilde{W})(\partial_{\theta}\tilde{W})
\right]
=0,
\end{equation}
while an independent combination of the diagonal equations yields
\begin{equation}\label{errethetatheta-main}
\mathcal{E}_{rr}-\mathcal{E}_{\theta\theta}
=
\eta\Bigg[
\Delta(\partial_{r}^{2}\tilde{W})
+(\partial_{\theta}^{2}\tilde{W})
+\frac{2r\Delta}{\Sigma}(\partial_{r}\tilde{W})
-\frac{2a^{2}\sin\theta\cos\theta}{\Sigma}
(\partial_{\theta}\tilde{W})
\end{equation}
\begin{equation*}
\qquad\qquad\qquad
-\frac{\Delta}{\Sigma^{2}}
\left(
\Delta(\partial_{r}\tilde{\Psi})^{2}
-(\partial_{\theta}\tilde{\Psi})^{2}
\right)
\Bigg]
=0.
\end{equation*}

Equations~\eqref{PDE-main}, \eqref{ertheta-main}, and
\eqref{errethetatheta-main} illustrate the nontrivial conditions that arise
once the scalar sector is generalized from the static ansatz to a stationary
and axisymmetric configuration. In particular, they show that the
inconsistency encountered with $\Phi=qt+\Psi(r)$ cannot, in general, be
removed by simply retaining the static scalar profile after the
Newman--Janis transformation. A scalar configuration capable of supporting
the rotating metric~\eqref{rle}, if such a regular configuration exists,
would therefore require a nontrivial angular dependence.

We stress, however, that deriving these conditions does not by itself
establish the existence of a regular scalar configuration supporting
the metric~\eqref{rle}. A complete demonstration would require solving
the relevant coupled field equations subject to appropriate boundary
conditions, verifying regularity at the event horizon and on the symmetry
axis, recovering the static scalar configuration in the limit $a\to0$,
imposing the appropriate asymptotic behavior, and checking the remaining
metric and scalar equations on the resulting configuration. Such an analysis
is beyond the scope of the present work.

Accordingly, in what follows, the metric~\eqref{rle} is treated as a
Newman--Janis-generated phenomenological rotating extension of the exact
static primary-scalar-hair geometry. Its horizon structure and null-geodesic
phenomenology can be investigated at the metric level, but the resulting
observational constraints should be interpreted as constraints on this
rotating metric parametrization rather than as direct constraints on an
established exact rotating solution of the underlying beyond-Horndeski
theory.

The rotating metric ~\eqref{rle} depends, in addition to the mass parameter $M$
and spin parameter $a$, on the parameters inherited from the static seed
solution, namely $q$, $\eta$, and $\lambda$, through the effective
combination $Q=\eta q^4$. Since $q^4\geq0$, the sign of $Q$ is fixed
by the sign of $\eta$, whereas its magnitude depends on both $|\eta|$
and $|q|$. In the limit $q\rightarrow0$ at fixed $\eta$, and hence
$Q\rightarrow0$, the deformation vanishes and the metric smoothly
reduces to the Kerr geometry. Conversely, in the static limit
$a\rightarrow0$, the metric reduces to the exact asymptotically flat
black-hole solution with primary scalar hair given by Eq.~\eqref{ssle}.

Similar to the Kerr black hole, the line element of the rotating black hole 
with primary scalar hair \eqref{rle} has two invariance isometries as 
time-translational and rotational symmetries corresponding with two Killing 
vectors $\chi^{\mu}_{(t)}=\left(\frac{\partial}{\partial t}\right)^{\mu}$ and 
$\chi^{\mu}_{(\varphi)}=\left(\frac{\partial}{\partial\varphi}\right)^{\mu}$, 
respectively. These Killing vectors admit two conserved quantities in 
determining the motion of the test particles (omitting the back reaction) 
around the rotating black hole: the total energy $E$ and the axial angular 
momentum $L_{z}$.

The ring singularity of the rotating black hole with primary scalar hair 
\eqref{rle} occurs at $\Sigma=0$. On the other hand, the horizons (coordinate 
singularities) of the rotating black hole occur at $r$ values satisfying 
$\Sigma\neq 0$ and $g^{\mu\nu}\partial_{\mu}r\partial_{\nu}r=g^{rr}=\Delta=0$ in 
which $g_{\mu\nu}$ is the metric tensor of the rotating black hole with primary 
scalar hair \eqref{rle}. Therefore, the roots of
\begin{equation}\label{hor}
r^{2}+a^{2}-2Mr+rQ\lambda\left[\frac{\pi}{2}+\frac{r\lambda}{\lambda^{2}+r^{2}}
-\arctan\left(\frac{r}{\lambda}\right)\right]=0\,,
\end{equation}
are the horizons of the rotating black hole with primary scalar hair 
\eqref{rle}. However, the analytical expressions of the horizons of the rotating 
black hole with primary scalar hair cannot be found due to complexity of Eq. 
\eqref{hor}. Hence, we proceed to find the horizons of the rotating black hole 
with primary scalar hair, numerically.

\begin{figure}[htb]
\centering
	\begin{tabular}{cc}
		\includegraphics[width=0.43\textwidth]{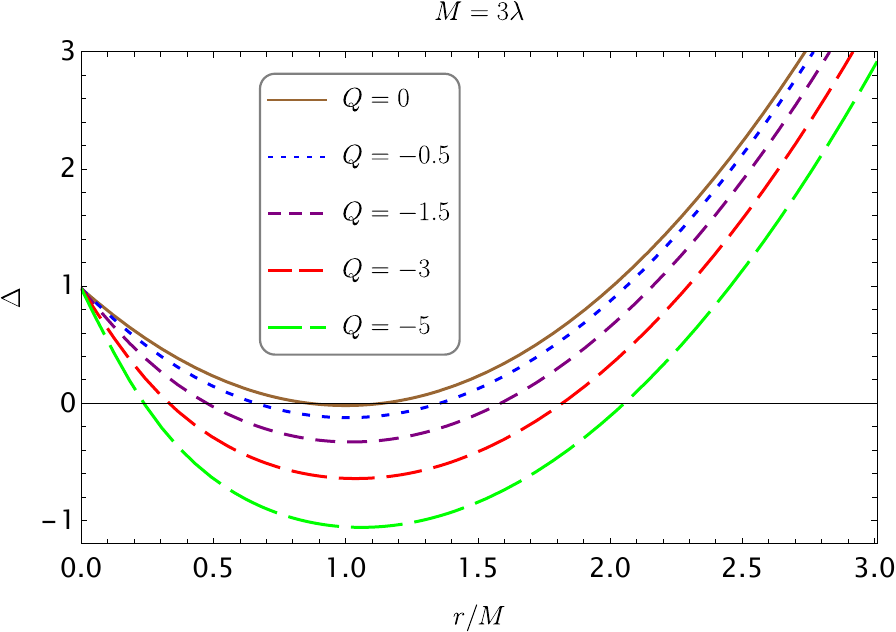}\,  
		\includegraphics[width=0.425\textwidth]{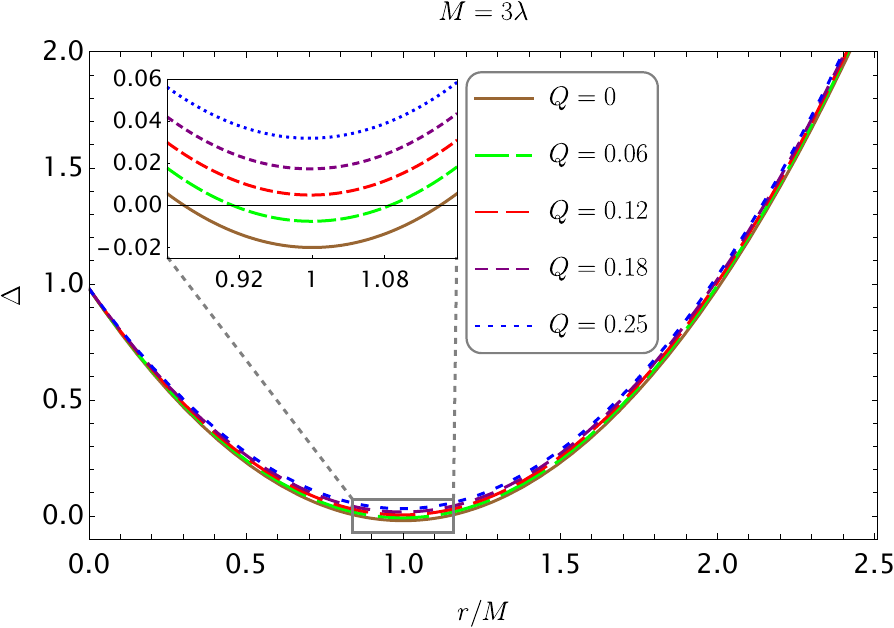}\, \\
		 \includegraphics[width=0.43\textwidth]{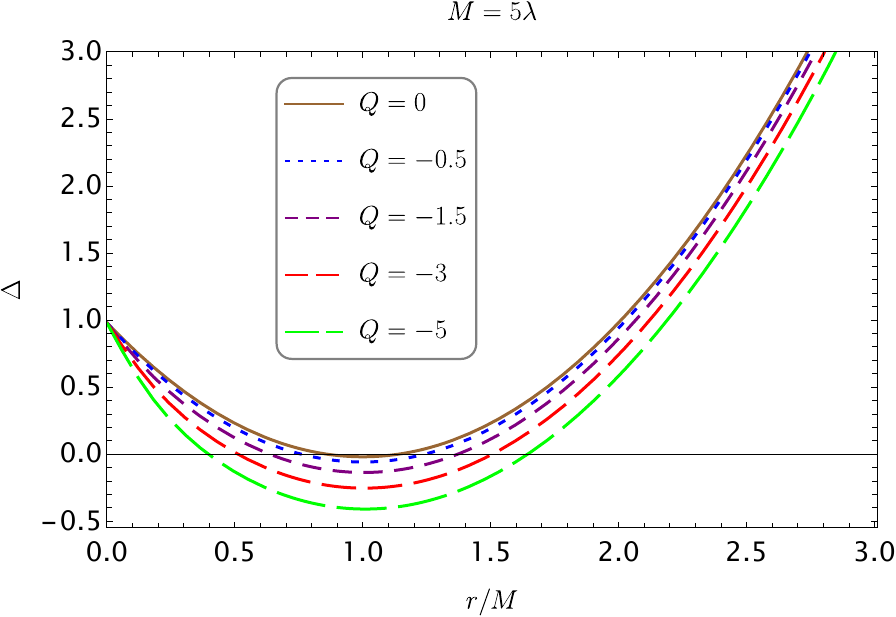}\,  
		\includegraphics[width=0.425\textwidth]{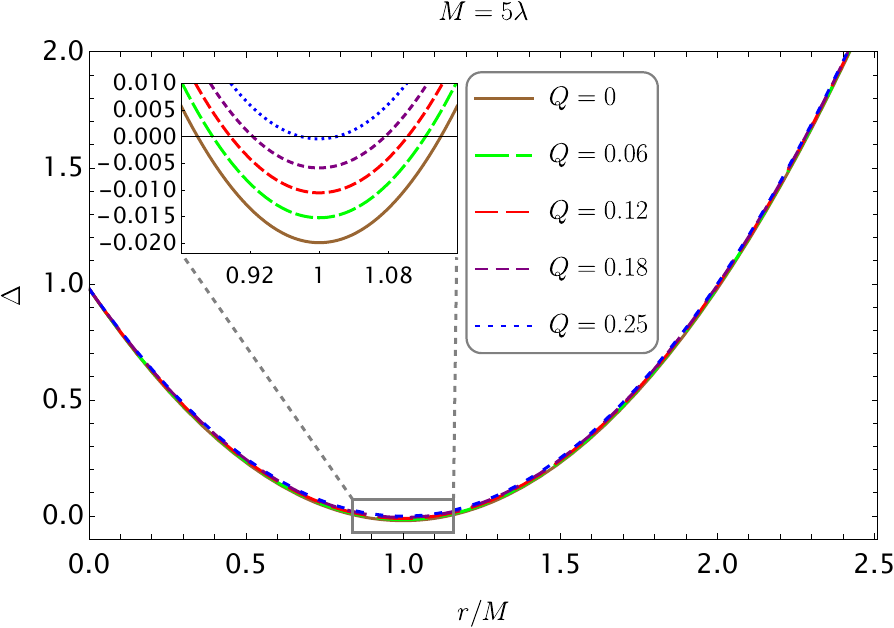}
	\end{tabular}
\caption{\label{Fig3}\small{\emph{The behavior of $\Delta$ versus $r$, for 
various values of $Q$ and $a=0.99$. The upper panels correspond to $M=3\lambda$ 
and the lower panels to $M=5\lambda$, while the left panels correspond to $Q<0$ 
and the right panels to $Q>0$.}}}
\end{figure}
In the upper panels of Fig.~\ref{Fig3}, we depict the behavior of $\Delta$ 
versus $r$ for $M=3\lambda$. As we can see, the black hole generally has two 
horizons: a Cauchy horizon $r_{\mathrm{C}}$ and an event horizon 
$\tilde{r}_{\mathrm{eh}}$. When $Q<0$ (i.e., $\eta<0$), the Cauchy horizon is 
always smaller than the Cauchy horizon of the Kerr black hole, whereas the event 
horizon is always larger than the Kerr event horizon. Additionally, increasing 
the magnitude $|Q|$ within the negative-$Q$ branch leads to a decrease of the 
Cauchy horizon and an increase of the event horizon of the rotating black hole 
with primary scalar hair. When $Q>0$ (i.e., $\eta>0$), the Cauchy horizon of 
the black hole is larger than the Cauchy horizon of the Kerr black hole, while 
its event horizon is smaller than the Kerr event horizon. In this situation, for 
$Q>0.06$, the rotating black hole with primary scalar hair is horizonless and 
becomes a naked singularity. Additionally, the lower panels of Fig.~\ref{Fig3} 
illustrate the behavior of $\Delta$ versus $r$ for $M=5\lambda$. We observe a 
similar behavior to that of the upper panels, except that for $Q>0$ 
(i.e., $\eta>0$) the rotating black hole with primary scalar hair is no longer 
horizonless.

Finally, for more transparency, in Table \ref{Table2}, we collect the values of 
the Cauchy horizon $r_{\mathrm{C}}$ and the event horizon 
$\tilde{r}_{\mathrm{eh}}$ of the rotating black hole with primary scalar hair in 
the parameter space.
\begin{table}[htb]
\centering
\caption{\label{Table2}\small{\emph{The values of $r_{\mathrm{C}}$ and 
$\tilde{r}_{\mathrm{eh}}$ of the rotating black hole with primary scalar hair in 
the parameter space. The case of $Q=0$ represents horizons of the Kerr black 
hole. We have set $a=0.99$.}}}
\begin{tabular}{c|c|cccc|cccc|c}
\multirow{2}{*}{horizons} & Kerr    & \multicolumn{4}{c|}{$Q<0$}                
                                                             & 
\multicolumn{4}{c|}{$Q>0$}                                                      
                         & \multirow{2}{*}{$\lambda$ values} \\ \cline{2-10}
                          & $Q=0$   & \multicolumn{1}{c|}{$Q=-0.5$} & 
\multicolumn{1}{c|}{$Q=-1.5$} & \multicolumn{1}{c|}{$Q=-3$}  & $Q=-5$  & 
\multicolumn{1}{c|}{$Q=0.06$} & \multicolumn{1}{c|}{$Q=0.12$} & 
\multicolumn{1}{c|}{$Q=0.18$} & $Q=0.25$ &                                   \\ 
\hline\hline
$r_{\mathrm{C}}$                   & $0.858$ & \multicolumn{1}{c|}{$0.659$}  & 
\multicolumn{1}{c|}{$0.474$}  & \multicolumn{1}{c|}{$0.331$} & $0.236$ & 
\multicolumn{1}{c|}{$0.912$}  & \multicolumn{1}{c|}{}         & 
\multicolumn{1}{c|}{}         &          & \multirow{2}{*}{$M=3\lambda$}     \\ 
\cline{1-10}
$\tilde{r}_{\mathrm{eh}}$          & $1.141$ & \multicolumn{1}{c|}{$1.356$}  & 
\multicolumn{1}{c|}{$1.586$}  & \multicolumn{1}{c|}{$1.819$} & $2.054$ & 
\multicolumn{1}{c|}{$1.085$}  & \multicolumn{1}{c|}{}         & 
\multicolumn{1}{c|}{}         &          &                                   \\ 
\hline
$r_{\mathrm{C}}$                   & $0.858$ & \multicolumn{1}{c|}{$0.758$}  & 
\multicolumn{1}{c|}{$0.635$}  & \multicolumn{1}{c|}{$0.511$} & $0.396$ & 
\multicolumn{1}{c|}{$0.876$}  & \multicolumn{1}{c|}{$0.897$}  & 
\multicolumn{1}{c|}{$0.922$}  & $0.979$  & \multirow{2}{*}{$M=5\lambda$}     \\ 
\cline{1-10}
$\tilde{r}_{\mathrm{eh}}$          & $1.141$ & \multicolumn{1}{c|}{$1.243$}  & 
\multicolumn{1}{c|}{$1.371$}  & \multicolumn{1}{c|}{$1.507$} & $1.644$ & 
\multicolumn{1}{c|}{$1.123$}  & \multicolumn{1}{c|}{$1.102$}  & 
\multicolumn{1}{c|}{$1.076$}  & $1.019$  &                                  
\end{tabular}
\end{table}

\begin{figure}[htb]
\centering
	\begin{tabular}{cc}
		\includegraphics[width=0.46\textwidth]{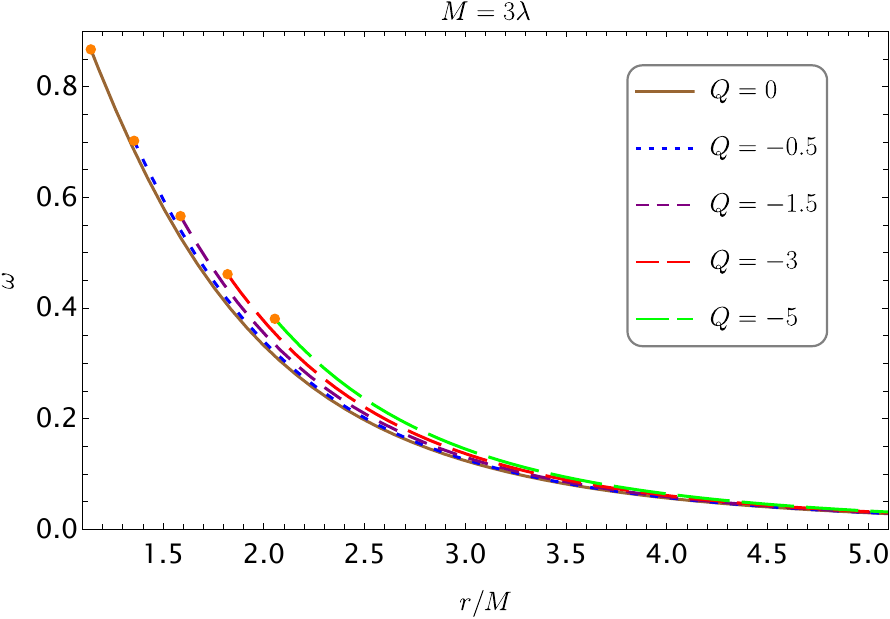}\, &
		\includegraphics[width=0.46\textwidth]{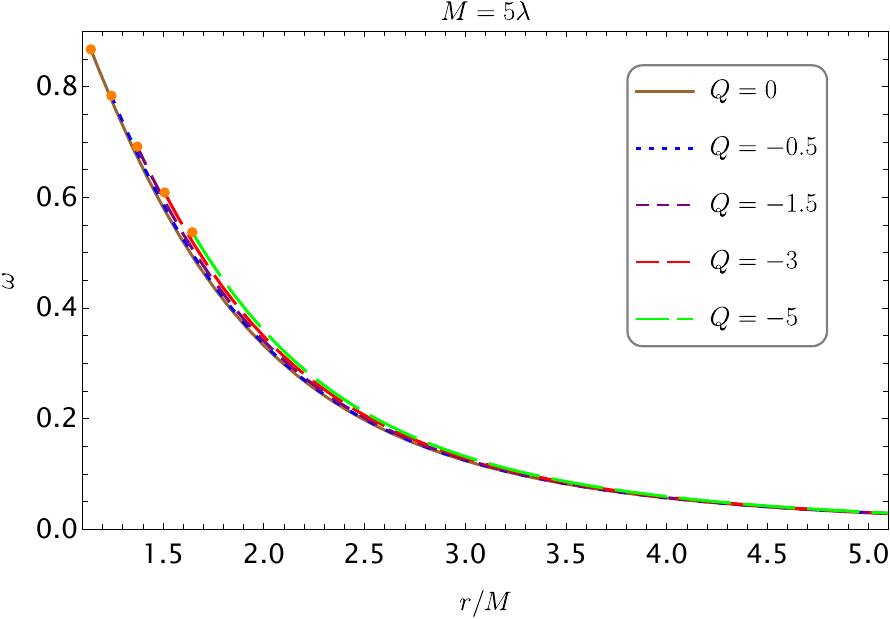}
	\end{tabular}
\caption{\label{omega}\small{\emph{
Angular velocity $\omega$  of the spacetime (frame dragging) as a function of 
the dimensionless radial coordinate   $r/M$, for different values of the scalar 
hair parameter $Q$, in a 
rotating scalar-hairy black hole configuration.}}}
\end{figure}

Stationary observers located outside the event horizon of a rotating black
hole, despite having zero angular momentum relative to an observer at
spatial infinity, are compelled to co-rotate due to the frame-dragging
effect. Their angular velocity is given by
\begin{equation}
\omega
=
\frac{d\phi}{dt}
=
-\frac{g_{t\phi}}{g_{\phi\phi}}.
\end{equation}
Since the mixed term in the line element is written as
$2g_{t\phi}\,dt\,d\phi$, Eq.~(8) gives
\begin{equation}
g_{t\phi}
=
-\frac{2a\rho(r)\sin^2\theta}{\Sigma}
=
-\frac{a\left(r^2+a^2-\Delta\right)\sin^2\theta}{\Sigma},
\end{equation}
where $2\rho(r)=r^2+a^2-\Delta$. Furthermore,
\begin{equation}
g_{\phi\phi}
=
\frac{\sin^2\theta}{\Sigma}
\left[
\left(r^2+a^2\right)^2
-a^2\Delta\sin^2\theta
\right].
\end{equation}
Hence, the frame-dragging angular velocity takes the form
\begin{equation}
\omega
=
\frac{
a\left(r^2+a^2-\Delta\right)
}{
\left(r^2+a^2\right)^2
-a^2\Delta\sin^2\theta
}.
\label{omega_correct}
\end{equation}
In Fig. \ref{omega}, we present the behavior of the angular velocity. As we can 
see, $\omega$ increases monotonically as $r$ decreases reaching its maximum 
value at $r=r_{\mathrm{eh}}$. At the event horizon, where
$\Delta(r_{\rm h})=0$, the frame-dragging angular velocity becomes
independent of $\theta$ and reduces to
\begin{equation}
\Omega_{\rm H}
=
\left.\omega\right|_{r=r_{\rm h}}
=
\frac{a}{r_{\rm h}^2+a^2}.
\label{OmegaH_correct}
\end{equation}
which corresponds to the black hole angular velocity $\Omega$. This expression reduces to the standard Kerr horizon angular velocity in
the limit $Q\rightarrow0$.
 
\begin{figure}[htb]
\centering
	\begin{tabular}{cc}
        \includegraphics[width=0.46\textwidth]{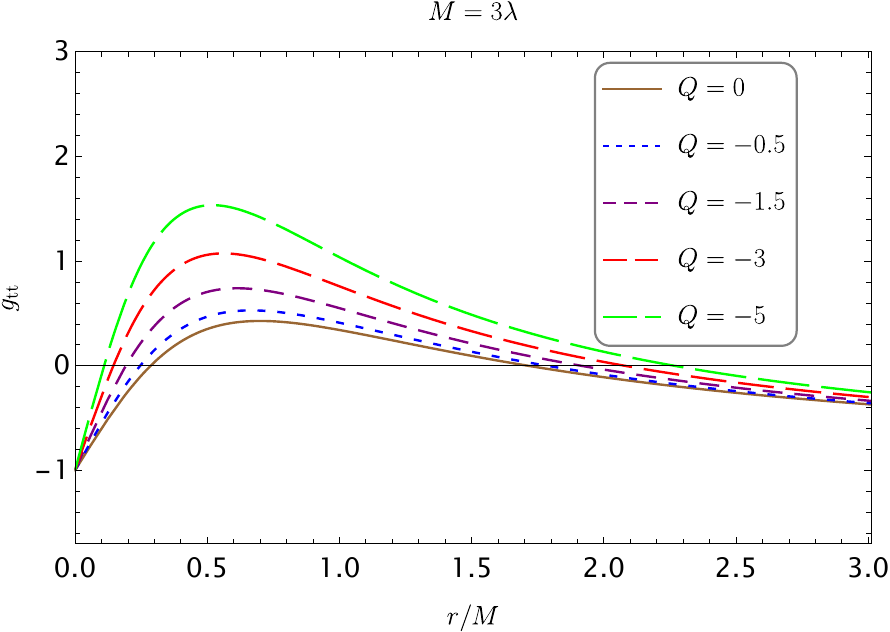}\, &
		\includegraphics[width=0.455\textwidth]{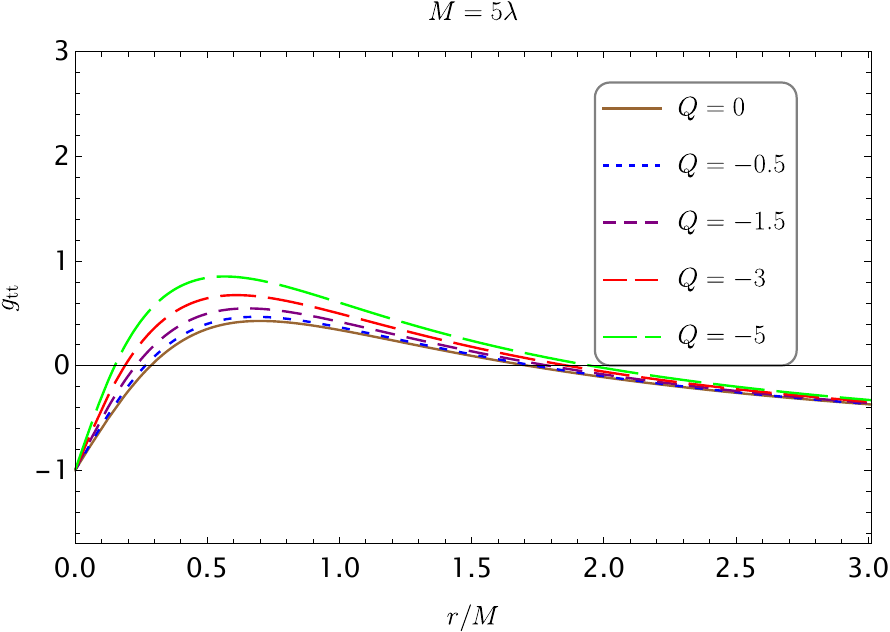}
	\end{tabular}
\caption{\label{Figgt}\small{\emph{The behavior of SLS versus $r$ for various 
values of $Q$ with $a=0.99$.}}}
\end{figure}

While stationary observers may exist anywhere outside the event horizon, static 
observers-whose trajectories are aligned with the timelike Killing vector 
$\eta^{\mu}_{(t)}$-are restricted to the regions outside the static limit 
surface (SLS), defined by the condition 
$\eta^{\mu}_{(t)}\eta_{\mu(t)}=g_{tt}=0$ \cite{Chandrasekhar:1985kt}. The 
locations of the SLS are obtained from the roots of the equation
\begin{equation}\label{sls}
r^{2}+a^{2}\cos^{2}\theta-2Mr+rQ\lambda\left(\frac{\pi}{2}+\frac{r\lambda}{
\lambda^{2}+r^{2}}-\arctan\left(\frac{r}{\lambda}\right)\right)=0\,,
\end{equation}
which, in addition to the black hole parameters, also depends on the polar 
angle $\theta$, and matches the event horizon only at the poles. We solve Eq. 
\eqref{sls} numerically, and the two positive real roots representing the two 
SLS are plotted in Fig. \ref{Figgt}, with $M=3\lambda$ and $M=5\lambda$ 
respectively, for different values of $Q$. We see that the radius of the outer 
SLS decreases as $Q$ increases, and moreover we observe that for fixed values 
of $M$ and $a$ the outer SLS radius of the rotating black hole with primary 
scalar hair are larger than that of a Kerr black hole.

\begin{figure}[htb]
    \centering
    \begin{subfigure}{0.3\textwidth}
        \includegraphics[width=\linewidth]{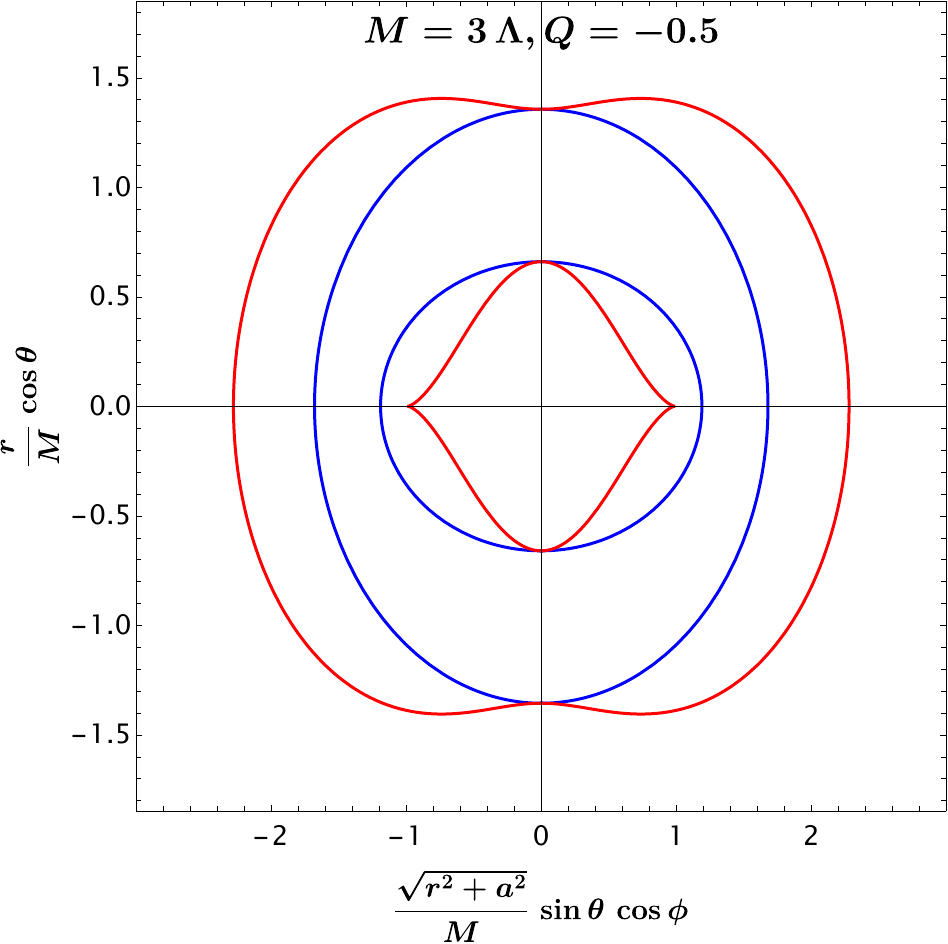}
    \end{subfigure}
    \begin{subfigure}{0.3\textwidth}
        \includegraphics[width=\linewidth]{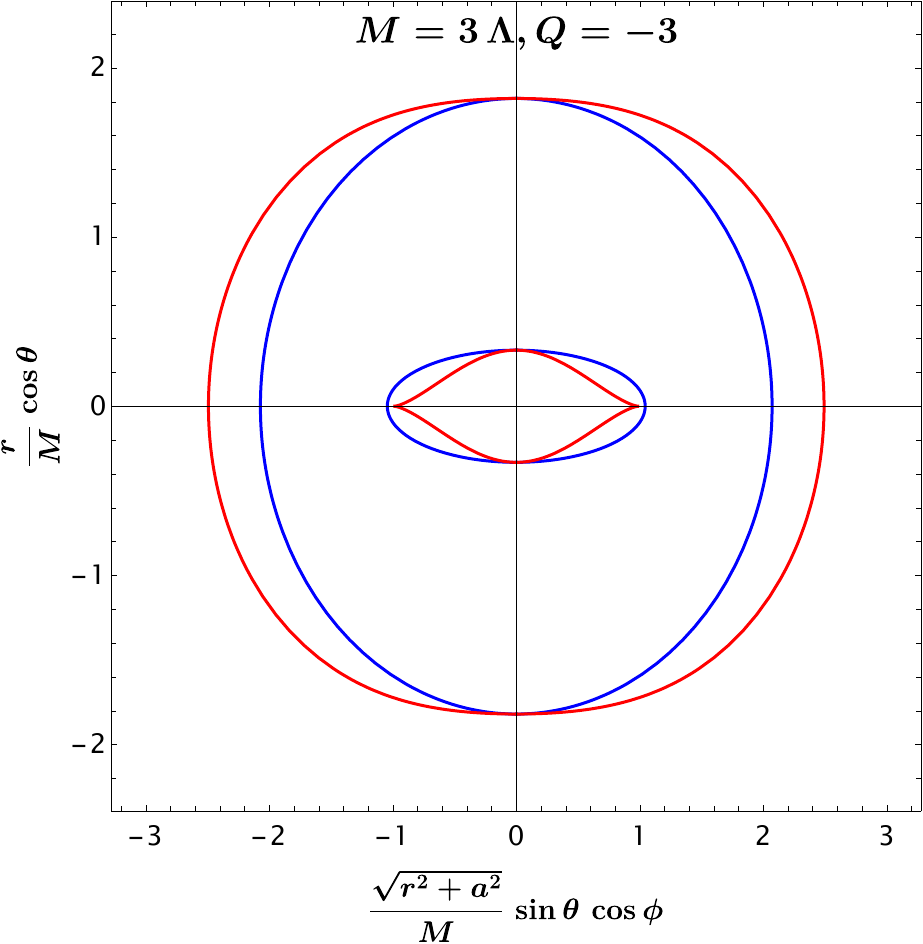}
    \end{subfigure}
    \begin{subfigure}{0.3\textwidth}
    \includegraphics[width=\linewidth]{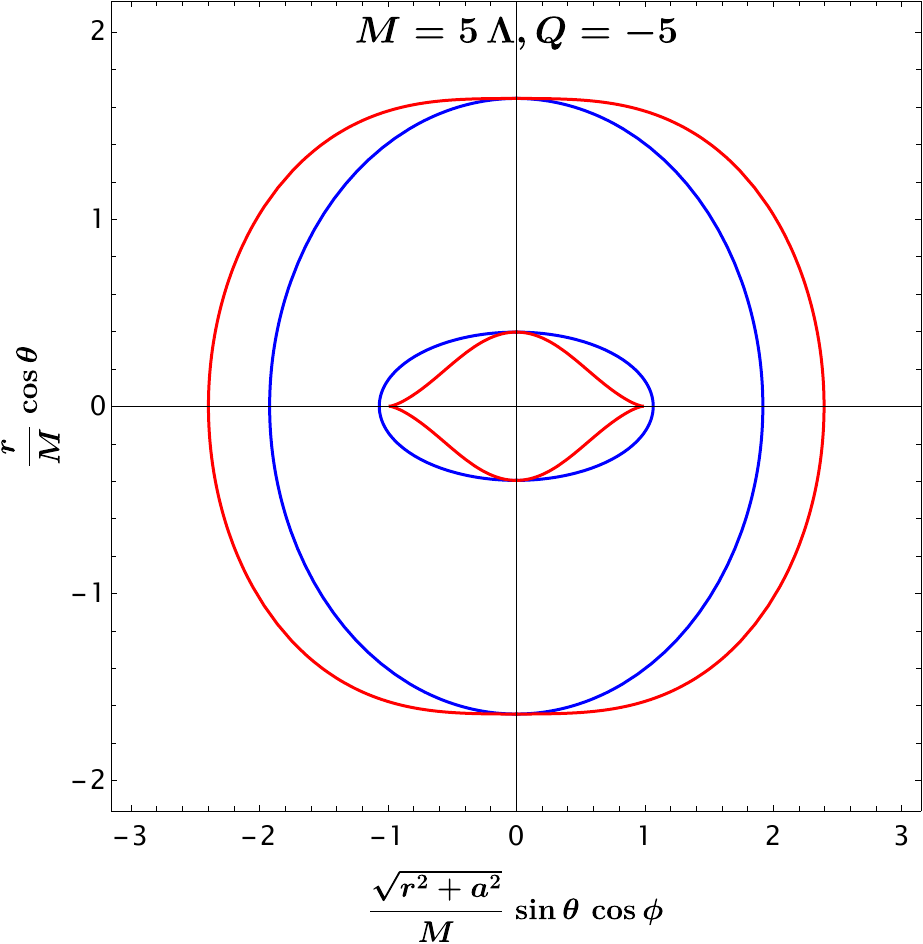}
    \end{subfigure}

    \begin{subfigure}{0.3\textwidth}
        \includegraphics[width=\linewidth]{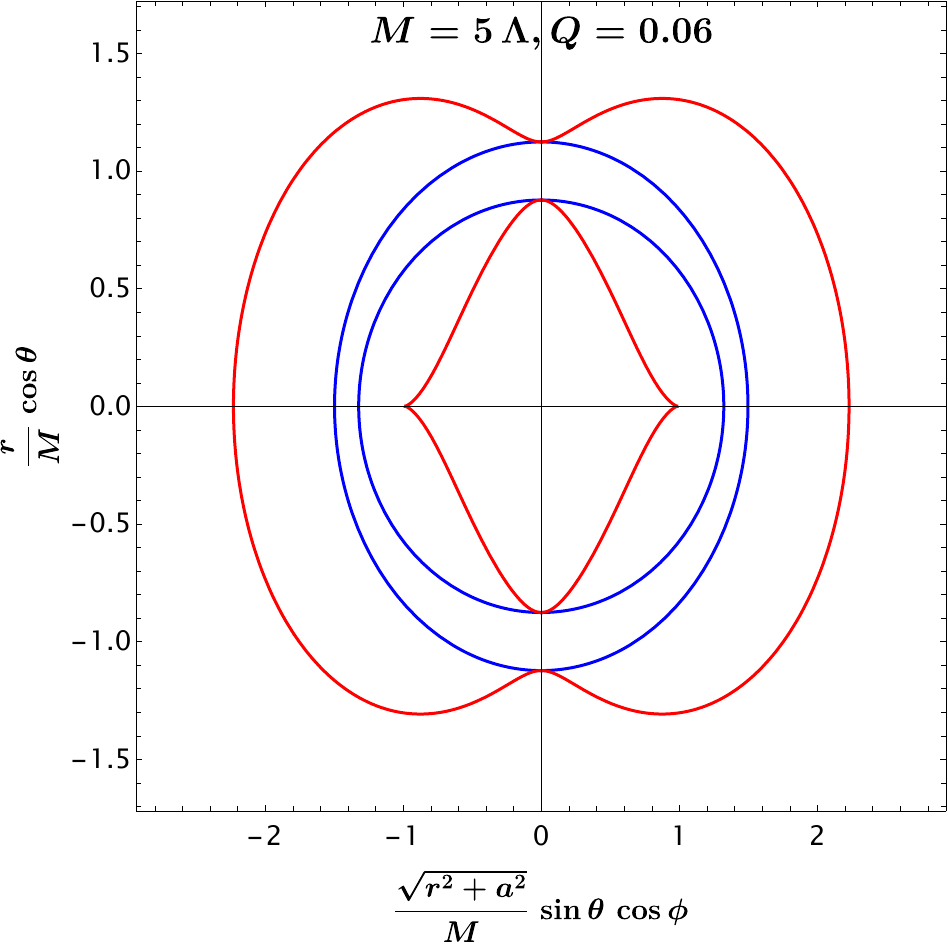}
    \end{subfigure}
    \begin{subfigure}{0.3\textwidth}
        \includegraphics[width=\linewidth]{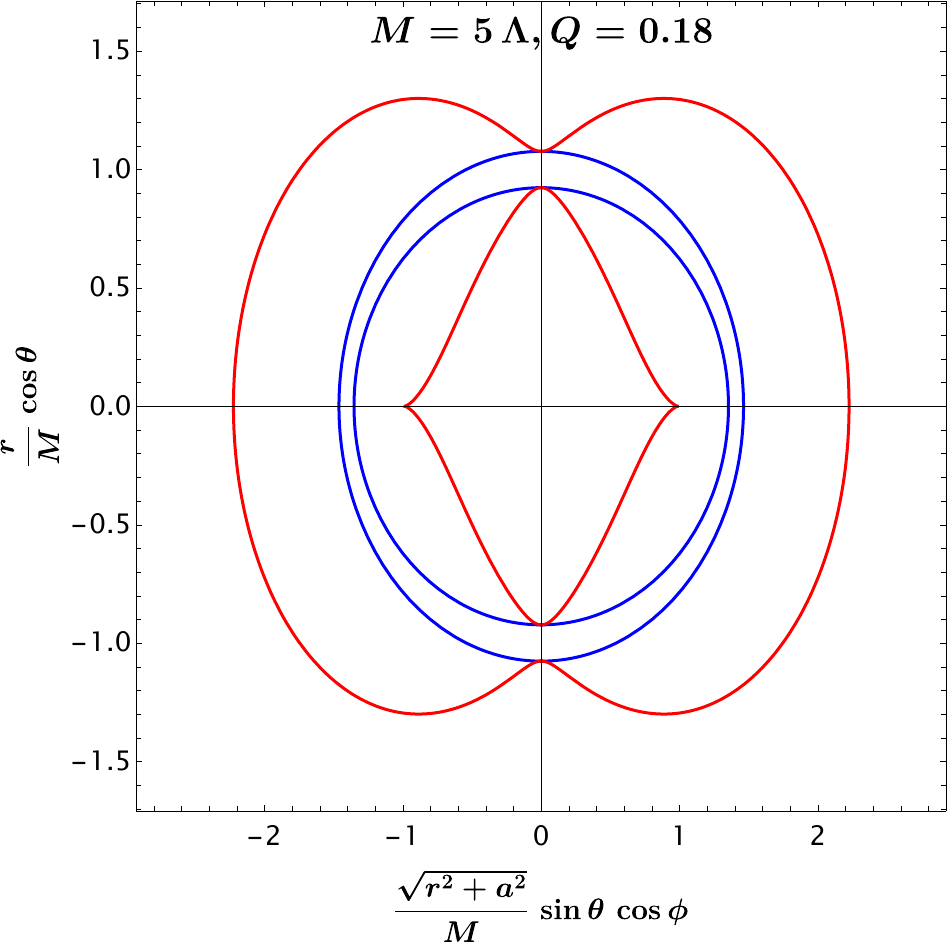}
    \end{subfigure}
    \begin{subfigure}{0.3\textwidth}
        \includegraphics[width=\linewidth]{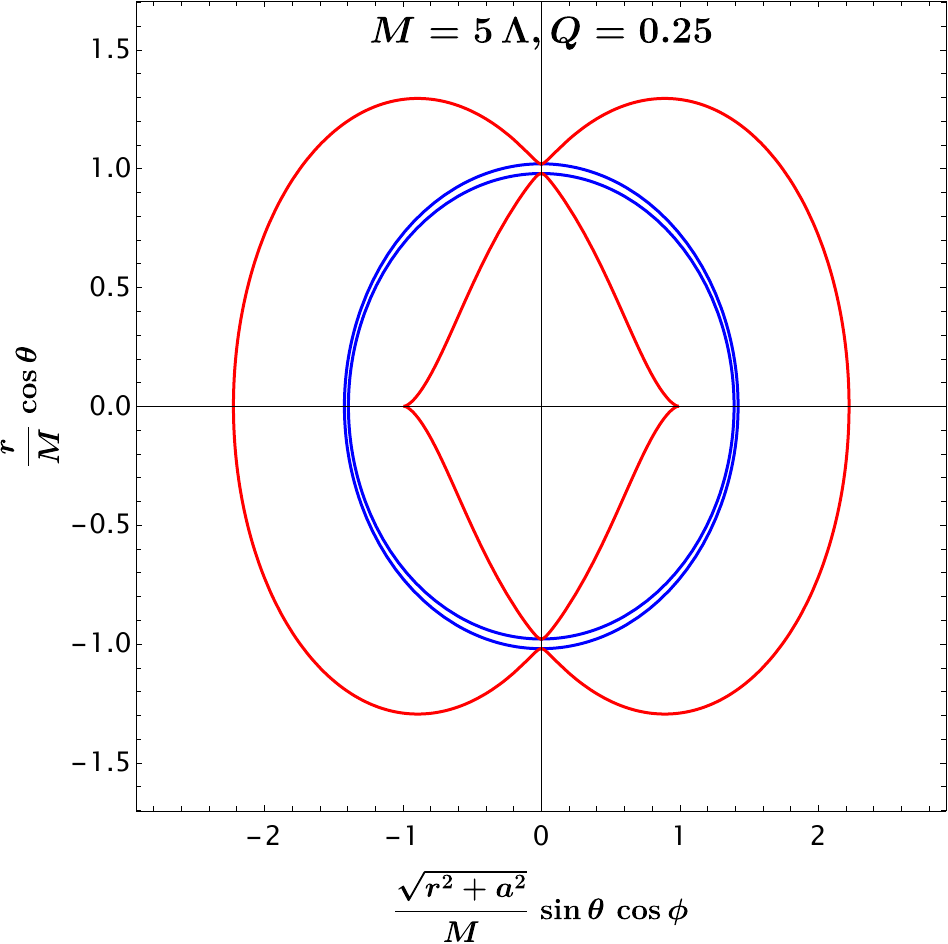}
    \end{subfigure}

    \begin{subfigure}{0.3\textwidth}
        \includegraphics[width=\linewidth]{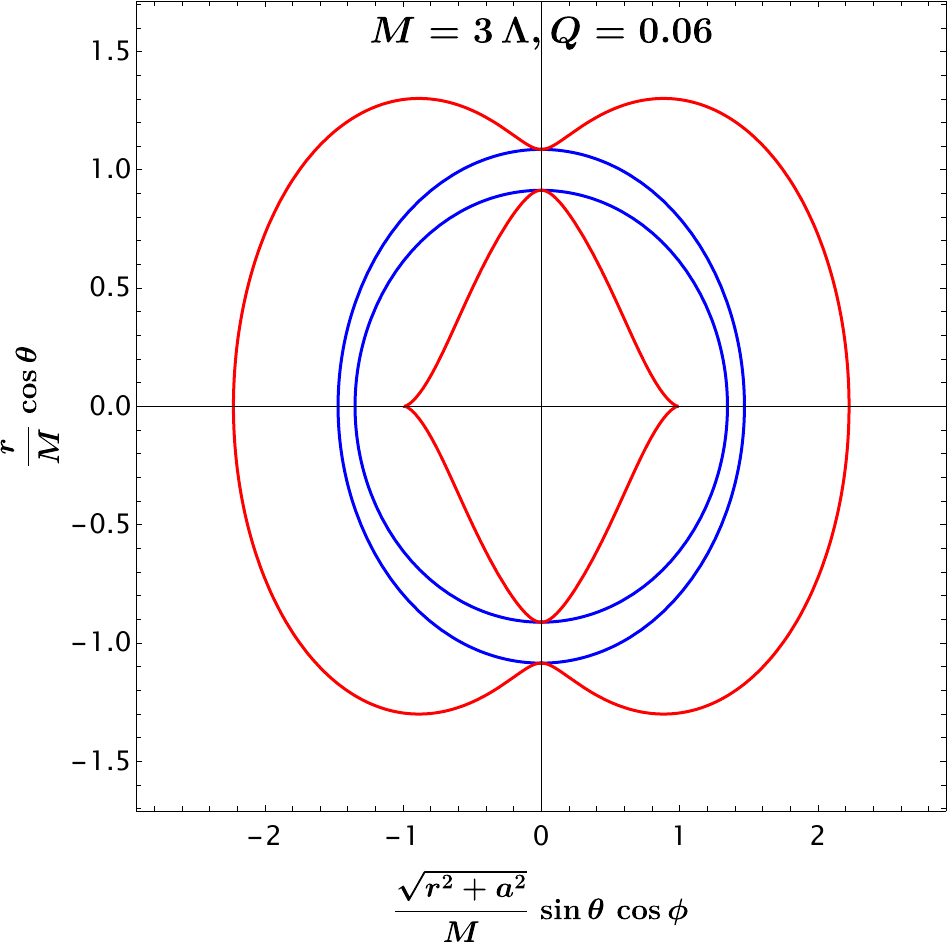}
    \end{subfigure}
    \begin{subfigure}{0.3\textwidth}
        \includegraphics[width=\linewidth]{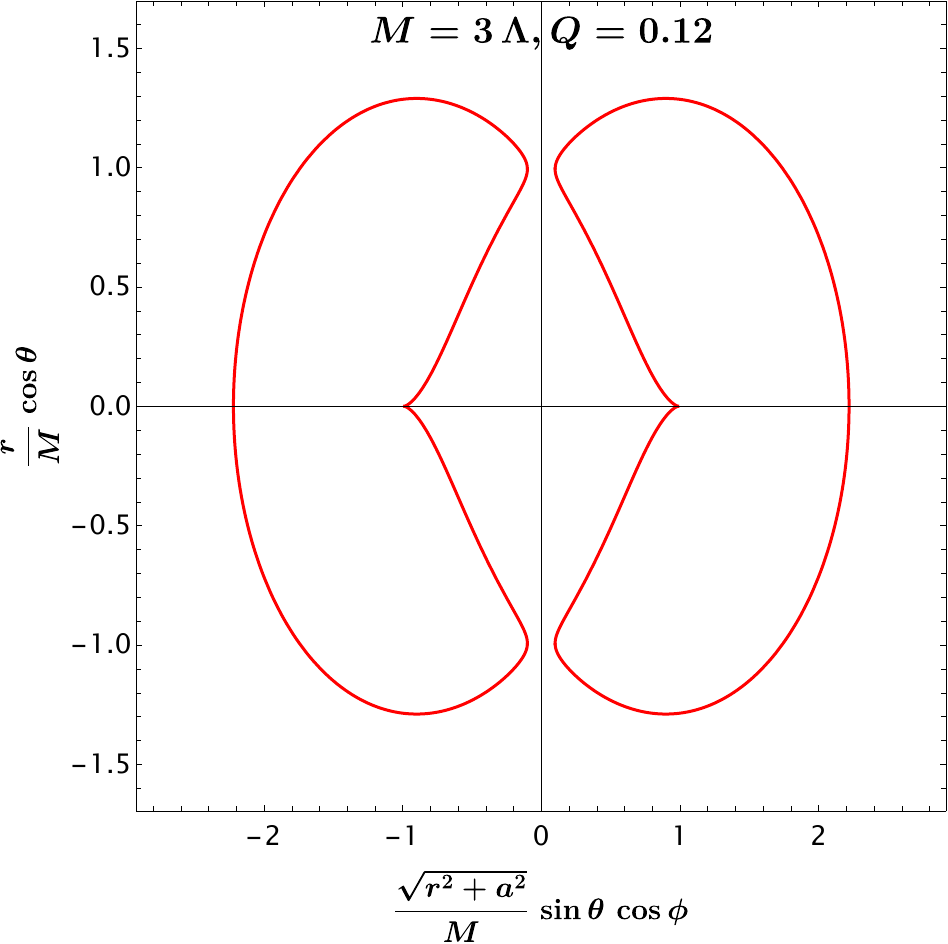}
    \end{subfigure}
    \begin{subfigure}{0.3\textwidth}
        \includegraphics[width=\linewidth]{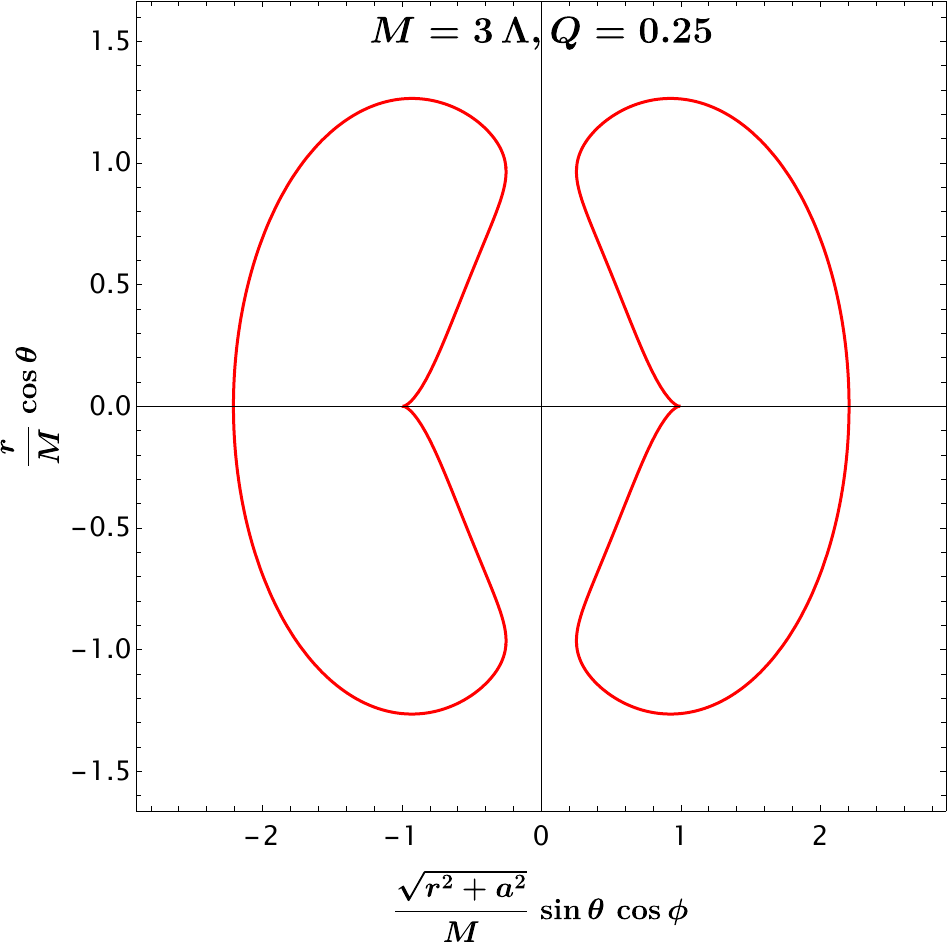}
    \end{subfigure}
    \caption{\label{fig:3x3}\small{\emph{
    Cross-sectional profiles of the event horizon (blue), the static limit 
surface (SLS) (red), and the enclosed ergoregion,  for various values of the 
parameters $\lambda$ and $Q$ with $a=0.99$.}}}
\end{figure}

The region located between the SLS and the event horizon is referred to as the 
ergoregion. It has been demonstrated that, at least theoretically through the 
Penrose process~\cite{Penrose:1969pc}, energy can be extracted from the black 
hole's ergosphere, which lies outside the event horizon. In Fig.~\ref{fig:3x3}, 
we show the ergoregion of rotating black hole with primary scalar hair. 
Interestingly, we find that the shape of the ergosphere becomes increasingly 
prolate as the scalar hair $Q$ grows. This indicates that rotating black holes 
in beyond Horndeski gravity with larger scalar hair values tend to have a more 
elongated ergosphere, which results in a larger ergospheric area. Furthermore, 
Fig.~\ref{fig:3x3} shows that there exists some values of the scalar hair for 
$M=3\lambda$, at which the two surfaces merge into a single one, i.e., for 
$Q>0.06$ the ergoregion disappears entirely. It would be worthwhile to 
investigate how the scalar hair affects the efficiency of energy extraction, 
however this study is left for a future project.

\section{Photon Regions and Shadow Formation}\label{sec:3}
 
Having introduced the phenomenological rotating metric \eqref{rle}, we now
investigate photon motion in this geometry. The properties of its null
geodesics determine the corresponding photon region and shadow. The analysis
in this section is therefore performed entirely at the metric level and does
not rely on assuming that \eqref{rle} has been established as an exact
rotating solution of the underlying beyond-Horndeski field equations.
 The properties of null geodesics determine the photon region and, 
consequently, the observable black-hole shadow. Since the shadow boundary 
corresponds to unstable spherical photon orbits, its morphology provides a 
direct probe of the underlying metric functions. In stationary and axisymmetric 
spacetimes, the geodesic structure is governed by conserved quantities 
associated with time-translation and rotational symmetries. In addition, 
separability of the Hamilton-Jacobi equation allows the introduction of a third 
constant of motion, leading to a complete characterization of photon 
trajectories. This structure enables a systematic determination of the spherical 
photon orbits and the associated critical impact parameters.

In the following subsections we first derive the null geodesic equations using 
the Hamilton-Jacobi formalism and then we construct the photon region and shadow 
observables. Particular attention is given to how the scalar hair parameter 
modifies the radial potential and, therefore, the location and stability of 
photon orbits relative to the Kerr limit.

\subsection{Null geodesics and photon orbits}

We now investigate null geodesics in the spacetime of a rotating black 
hole configuration with primary scalar hair in beyond Horndeski gravity, 
described by the line element \eqref{rle}. Since the metric is stationary and 
axisymmetric, photon motion admits conserved quantities associated with the 
Killing vectors $\partial_{t}$ and $\partial_{\varphi}$. A systematic way to 
derive the equations of motion is through the Hamilton-Jacobi formalism.

The Hamilton-Jacobi equation reads
\begin{equation}\label{hje}
\frac{\partial\mathcal{S}}{\partial\tau}+\mathcal{H}=0\,,
\end{equation}
where $\mathcal{S}$ is the Jacobi action, depending on the spacetime coordinates 
$x^{\mu}$ and the affine parameter $\tau$. The Hamiltonian for photon motion is
\begin{equation}\label{hpm}
\mathcal{H}=\frac{1}{2}g^{\mu\nu}p_{\mu}p_{\nu}\,,
\end{equation}
with conjugate momentum components
\begin{equation}\label{ccmc}
p_{\mu}=\frac{\partial \mathcal{S}}{\partial 
x^{\mu}}=g_{\mu\nu}\frac{\mathrm{d}x^{\nu}}{\mathrm{d}\tau}\,.
\end{equation}
Since the metric \eqref{rle} does not depend explicitly on $t$ and $\varphi$, 
the corresponding conjugate momenta are conserved. We therefore define 
$p_{t}=-E$ and $p_{\varphi}=L_{z}$, where $E$ and $L_{z}$ denote the photon 
energy and axial angular momentum, respectively.

Assuming separability of the Jacobi action, we write
\begin{equation}\label{ja}
\mathcal{S}=\frac{1}{2}m^{2}\tau-Et+L_{z}\varphi+\mathcal{S}_{r}(r)+\mathcal{S}_
{\theta}(\theta)\,,
\end{equation}
where $m^{2}=-p_{\mu}p^{\mu}$ is the mass of the test particle. In the present 
analysis we consider photons, hence $m=0$. The functions $\mathcal{S}_{r}(r)$ 
and $\mathcal{S}_{\theta}(\theta)$ are determined by inserting Eq.~\eqref{ja} 
into the Hamilton-Jacobi equation \eqref{hje}. After straightforward algebraic 
manipulations, we obtain
\begin{equation}\label{shje}
-\Delta\left(\frac{\mathrm{d}\mathcal{S}_{r}(r)}{\mathrm{d}r}\right)^{2}+\frac{
\left[\left(a^{2}+r^{2}\right)E-aL_{z}\right]^{2}}{\Delta}-(aE-L_{z})^{2}
=\left(\frac{\mathrm{d}\mathcal{S}_{\theta}(\theta)}{\mathrm{d}\theta}\right)^{2
}+\left(\frac{L_{z}^{2}}{\sin^{2}\theta}-a^{2}E^{2}\right)\cos^{2}\theta\,.
\end{equation}
The left-hand side of Eq.~(28) depends only on $r$, whereas the
right-hand side depends only on $\theta$. Therefore, both sides must be
equal to the same separation constant, which we denote by
$\mathcal{K}$. We reserve the symbol $Q$ exclusively for the scalar-hair
parameter introduced in Eq.~(5), and use $\mathcal{K}$ throughout the
geodesic analysis for the Carter-type separation constant. Accordingly, Eq.~(28) can be separated as
\begin{equation}\label{a}
\mathcal{K}
=
-\Delta
\left[
\frac{dS_r(r)}{dr}
\right]^2
+
\frac{
\left[
(r^2+a^2)E-aL_z
\right]^2
}{\Delta}
-
(aE-L_z)^2,
\end{equation}
and
\begin{equation}\label{b}
\mathcal{K}
=
\left[
\frac{dS_\theta(\theta)}{d\theta}
\right]^2
+
\left(
\frac{L_z^2}{\sin^2\theta}
-a^2E^2
\right)\cos^2\theta .
\end{equation}

From Eqs.~\eqref{a} and \eqref{b} we obtain
\begin{equation}\label{dsrdr}
\Delta\left[\frac{\mathrm{d}\mathcal{S}_{r}(r)}{\mathrm{d}r}\right]^{2}=\frac{
\left[\left(a^{2}+r^{2}\right)E-aL_{z}\right]^{2}-\Delta\left((aE-L_{z})^{2}
+\mathcal{K}\right)}{\Delta}\,,
\end{equation}
and
\begin{equation}\label{dsthdth}
\left(\frac{\mathrm{d}\mathcal{S}_{\theta}(\theta)}{\mathrm{d}\theta}\right)^{2}
=\mathcal{K}-\left(\frac{L_{z}^{2}}{\sin^{2}\theta}-a^{2}E^{2}\right)\cos^{2}
\theta\,,
\end{equation}
which determine the radial and angular parts of the motion.

Using Eq.~\eqref{ccmc}, we finally derive the geodesic equations for photons in 
the rotating black hole spacetime
\begin{equation}\label{dtdtau}
\Sigma\frac{\mathrm{d}t}{\mathrm{d}\tau}=\frac{\left(a^{2}+r^{2}\right)}{\Delta}
\left[\left(a^{2}+r^{2}\right)E-aL_{z}\right]-a\left(aE\sin^{2}\theta-L_{z}
\right)\,,
\end{equation}
\begin{equation}\label{drdtau}
\Sigma\frac{\mathrm{d}r}{\mathrm{d}\tau}=\pm\sqrt{\mathcal{R}(r)}\,,
\end{equation}
\begin{equation}\label{dthetadtau}
\Sigma\frac{\mathrm{d}\theta}{\mathrm{d}\tau}=\pm\sqrt{\Theta(\theta)}\,,
\end{equation}
and
\begin{equation}\label{dphidtau}
\Sigma\frac{\mathrm{d}\varphi}{\mathrm{d}\tau}=\frac{a}{\Delta}\left[\left(a^{2}
+r^{2}\right)E-aL_{z}\right]-\left(aE-\frac{L_{z}}{\sin^{2}\theta}\right)\,,
\end{equation}
where we have defined
\begin{equation}\label{RrThethth}
\mathcal{R}(r)\equiv\left[\left(a^{2}+r^{2}\right)E-aL_{z}\right]^{2}
-\Delta\left[(aE-L_{z})^{2}+\mathcal{K}\right]\,,
\qquad\Theta(\theta)\equiv\mathcal{K}-\left(\frac{L_{z}^{2}}{\sin^{2}\theta}-a^{
2}E^{2}\right)\cos^{2}\theta\,.
\end{equation}
The allowed region for photon motion is determined by the conditions 
$\mathcal{R}(r)\ge 0$ and $\Theta(\theta)\ge 0$, which ensure real-valued radial 
and angular trajectories.

\subsection{Shadow construction and observables}

In order to study the shadow of the  rotating scalar-hairy black hole 
configuration in beyond Horndeski gravity \eqref{rle}, we first define two 
impact factors, i.e.,
\begin{equation}\label{if}
\xi\equiv\frac{L_{z}}{E}\,,\qquad \zeta\equiv\frac{\mathcal{K}}{E^{2}}\,.
\end{equation}
With this notation, $Q$ refers exclusively to the scalar-hair deformation
parameter of the spacetime, whereas $\mathcal{K}$ and its normalized
counterpart $\zeta=\mathcal{K}/E^2$ refer exclusively to the conserved
separation constant of the photon motion.
Now, we can rewrite the functions $\mathcal{R}(r)$ and $\Theta(\theta)$ of Eq. 
\eqref{RrThethth} in terms of the aforementioned impact factors as  
\begin{equation}\label{RrTheththif}
\frac{\mathcal{R}(r)}{E^{2}}=\left(\left(a^{2}+r^{2}\right)-a\xi\right)^{2}
-\Delta\left((a-\xi)^{2}+\zeta\right)\,,\qquad 
\frac{\Theta(\theta)}{E^{2}}=\zeta-\left(\frac{\xi^{2}}{\sin^{2}\theta}-a^{2}
\right)\cos^{2}\theta\,,
\end{equation}
which again we require $\mathcal{R}(r)\geq 0$ and $\Theta(\theta)\geq 0$ as the 
allowed region for photon motion.

We aim to study the unstable photon orbits in the spacetime of the rotating 
black hole with primary scalar hair. According to the values of the critical 
impact parameters defined above, the photon rays originating from a light source 
can experience three possible situations: get captured into the rotating black 
hole; scatter to infinity; construct bound orbits around the black hole.

In studying the shadow of rotating black holes, the spherical (bound) null 
(photon) orbits living on a sphere with radius 
$r=\mathrm{const.}=r_{\mathrm{ph}}$ are important. Thus, these unstable 
spherical orbits have neither radial velocity $\dot{r}$ nor radial acceleration 
$\ddot{r}$, where an over dot stands for derivative with respect to $\tau$. 
Therefore, according to Eq. \eqref{drdtau}, the local extremum $r_{\mathrm{ph}}$ 
of the function $\mathcal{R}(r)$ is the radius of these unstable spherical 
photon orbits such that
\begin{equation}\label{cnupo}
\begin{split}
\mathcal{R}(r)\big|_{r=r_{\mathrm{ph}}}=0\,,\qquad\frac{\mathrm{d}\mathcal{R}(r)
}{\mathrm{d}r}\bigg|_{r=r_{\mathrm{ph}}}=0\,.
\end{split}
\end{equation}
Thus, solving simultaneously the set of Eqs. \eqref{cnupo} leads to the critical 
values of the impact parameters for unstable spherical photon orbits as follows
\begin{equation}\label{xietacrits}
\begin{split}
& 
\xi_{\mathrm{crit}}=\frac{\left(a^{2}+r^{2}\right)\Delta'-4r\Delta}{a\Delta'}\,,
\\
& 
\zeta_{\mathrm{crit}}=\frac{r^{2}\left(8\Delta\left(2a^{2}+r\Delta'\right)-r^{2}
\Delta'^{2}-16\Delta^{2}\right)}{a^{2}\Delta'^{2}}\,.
\end{split}
\end{equation}
It is worth noting that the unstable photon orbits are the boundary between 
capturing and scattering light rays by the rotating black hole, and therefore 
they are a fruitful tool for characterizing the shape of the shadow of the 
rotating black hole.

\begin{figure}[htb]
    \centering
    \begin{subfigure}{0.34\textwidth}
        \centering
        \includegraphics[width=\linewidth]{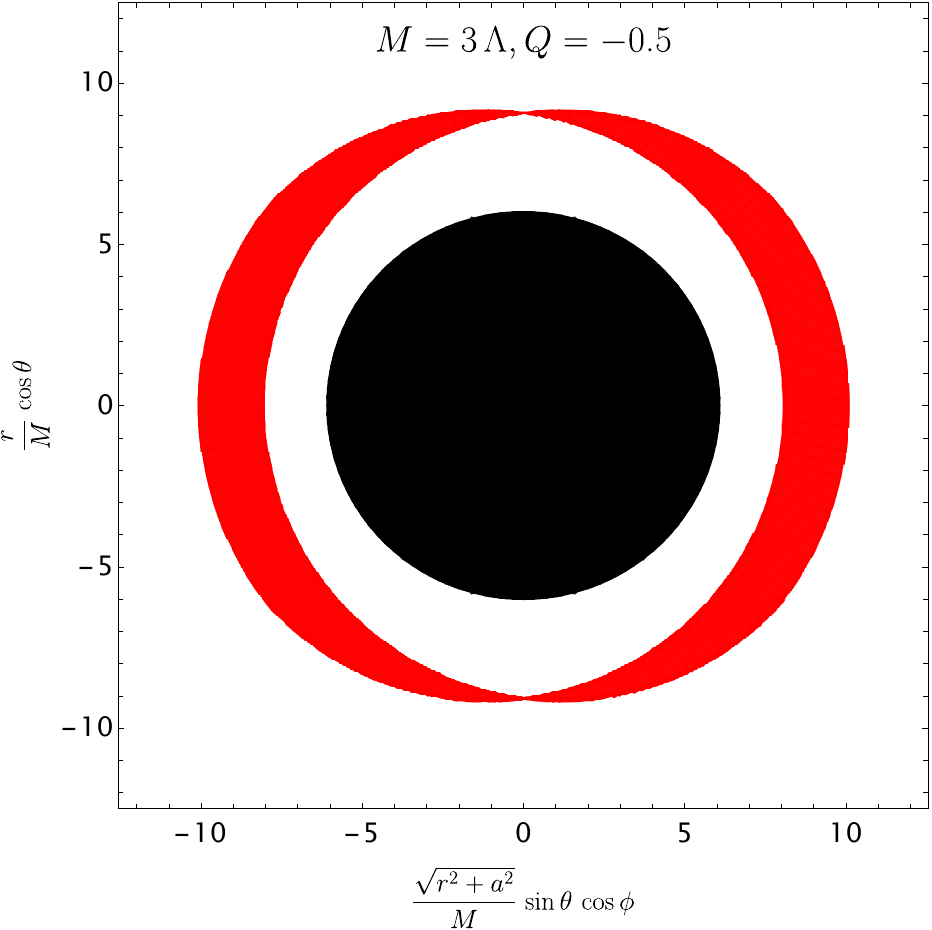}
    \end{subfigure}
    \begin{subfigure}{0.34\textwidth}
        \centering
        \includegraphics[width=\linewidth]{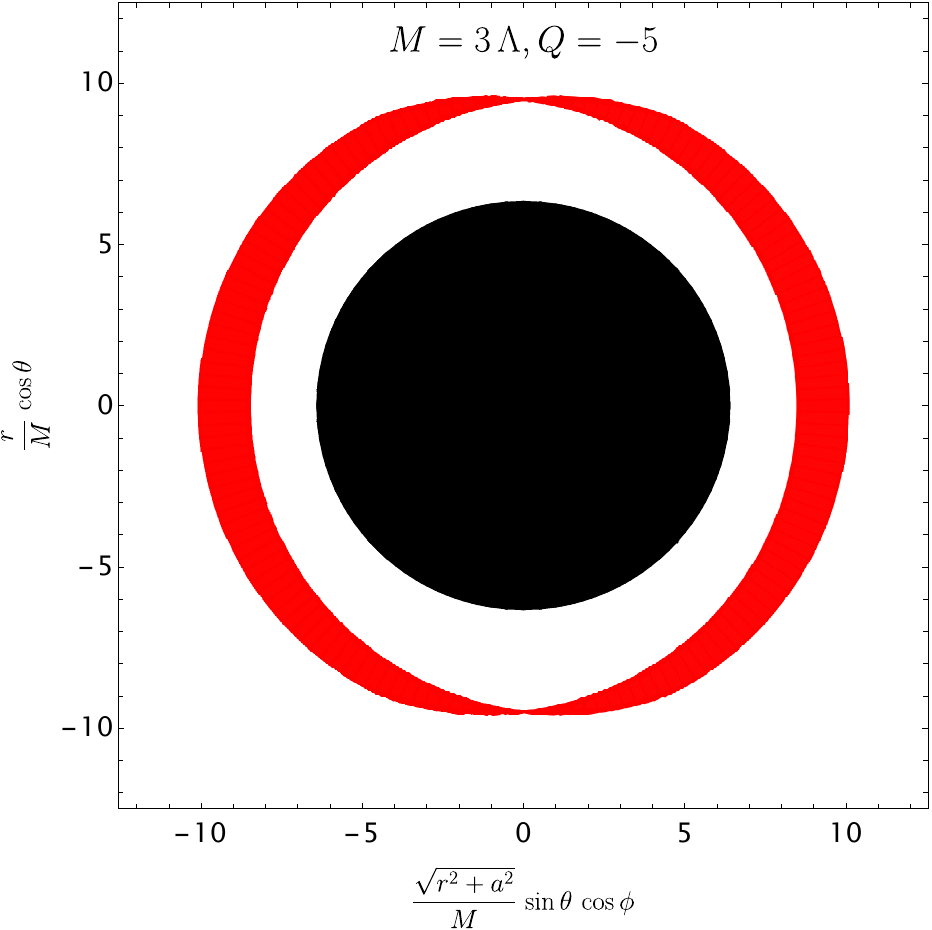}
    \end{subfigure}
    
    \begin{subfigure}{0.34\textwidth}
        \centering
        \includegraphics[width=\linewidth]{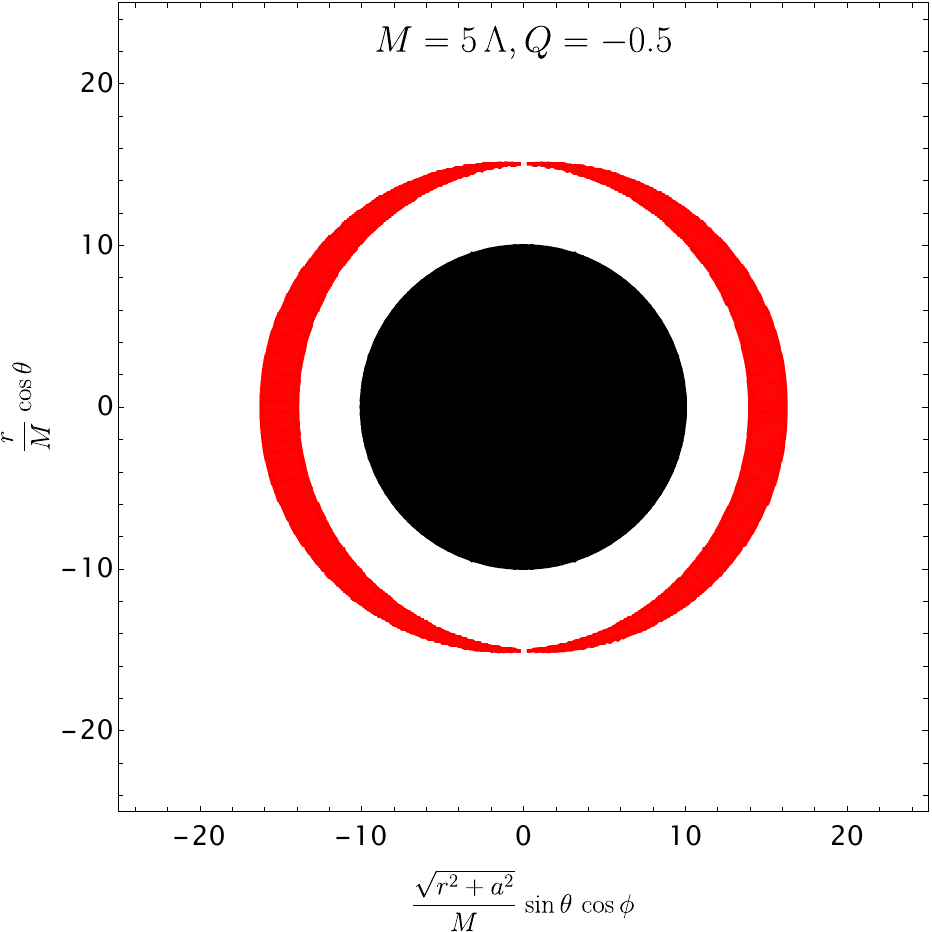}
    \end{subfigure}
    \begin{subfigure}{0.34\textwidth}
        \centering
        \includegraphics[width=\linewidth]{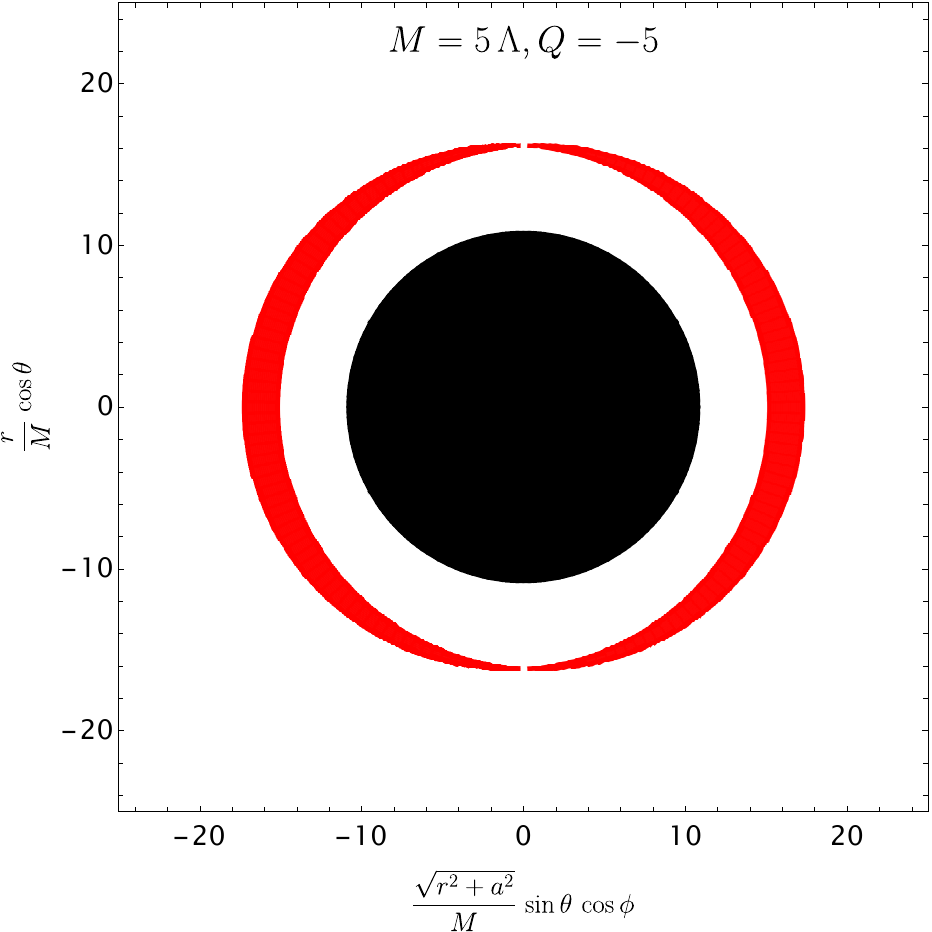}
    \end{subfigure}
    \caption{\label{fig:sixplots}\small{\emph{
    The configuration of the photon region (shaded red area), corresponding to 
unstable photon orbits, surrounding the rotating scalar-hairy black hole 
configuration (shaded black disk), for  $a=0.99$.}}}
\end{figure}

A crucial fact about rotating axisymmetric black hole spacetimes is that around 
them, there are prograde $r_{\mathrm{ph}}^{-}$ and retrograde 
$r_{\mathrm{ph}}^{+}$ unstable photon orbits on the equatorial plane. The 
prograde/retrograde unstable photon orbits are circulating the rotating black 
hole in the same/opposite direction as its rotation. Moreover, the prograde 
unstable photon orbits are smaller than the retrograde ones 
\cite{Bardeen:1975zz}. Furthermore, the generic spherical (3D and non-planar) 
photon orbits exist for $\zeta_{\mathrm{crit}}>0$ while the circular (planar) 
photon orbits can occur exclusively on the equatorial plane when 
$\zeta_{\mathrm{crit}}=0$. The radii $r_{\mathrm{ph}}^{\pm}$ of the co-rotating 
and counter-rotating circular photon orbits correspond to the real positive 
solutions of $\zeta_{\mathrm{crit}} = 0$. These spherical photon orbits form the 
photon region, which is determined by the equations \eqref{RrThethth} and 
\eqref{xietacrits}, subject to the condition $\Theta(\theta)\ge 0$. This is 
given by the following inequality
\begin{equation}
\left[4r_{p}\Delta(r_{p})-\Delta'(r_{p})\Sigma\right]^{2}\le 
16a^{2}r_{p}^{2}\Delta(r_{p})\sin^{2}\theta\,.
\end{equation}
As we can see in Fig. \ref{fig:sixplots}, the image produced by gravitational 
lensing of this photon region corresponds to the black hole shadow. Moreover, 
the photon region, formed by unstable photon orbits, varies with the scalar hair 
parameter $Q$ and the mass parameter $M$. The different panels show the 
configurations for various values of $Q$ (ranging from negative to positive), 
with $M=3\lambda$ and $M=5\lambda$, respectively. These plots highlight how the 
photon region's shape and size are influenced by changes in $Q$, providing 
insight into how scalar hair affects the black hole's gravitational field and 
photon dynamics.

The shadow of a black hole is the result of intense gravitational lensing near 
these compact objects, which distorts the light path and creates a dark zone 
visible to a distant observer 
\citep{Virbhadra:1999nm,Bozza:2001xd,Ghosh:2020spb,Kumar:2020owy,Islam:2020xmy}. 
This optical feature appears as a 2D shadow and is influenced by the geometry of 
the black hole's spacetime. The size and shape of the shadow can be used to 
determine key properties of the black hole, such as its mass, spin, and other 
characteristics. The shadow thus serves as an effective tool for testing 
modified gravity theories in strong gravitational fields 
\citep{Gott:2018ocn,Kumar:2020owy}, as well as the validity of the no-hair 
theorem \citep{Cunha:2016bjh}. The photon region surrounding the black hole's 
event horizon is composed of unstable photon orbits, marking the boundary 
between those photons that escape to infinity and those that fall into the black 
hole.

\begin{figure}[htb]
\centering
\includegraphics[width=0.4\textwidth]{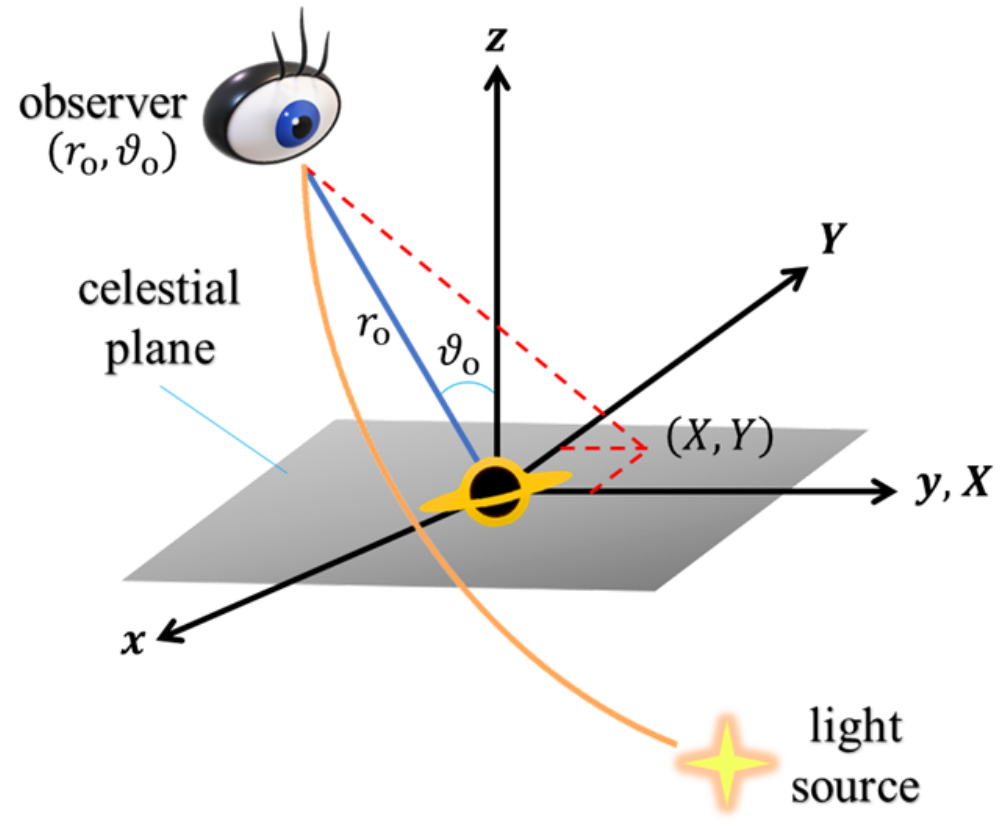}
\caption{\label{Fig5}\small{\emph{
Schematic diagram of the celestial coordinates on the observer’s sky. The 
quantity  $r_{\mathrm{o}}$ denotes the distance of the observer from the black 
hole, while $\vartheta_{\mathrm{o}}$  is the inclination angle. The coordinates 
$(X, Y)$ represent the apparent position of the shadow on the celestial plane. 
}}}
\end{figure}

We consider light sources distributed uniformly at infinity, with photons 
arriving with various impact parameters. These photons either scatter near the 
black hole's vicinity, or are captured by it. A distant observer is assumed to 
be at an inclination angle $\vartheta_{\mathrm{o}}$ relative to the black hole's 
rotation axis. The coordinates ($X$,$Y$) represent the angular distances of the 
shadow's boundary from the observer’s line of sight in directions perpendicular 
and parallel to the projected axis of rotation of the black hole onto the 
celestial plane \citep{Hioki:2009na}. By performing a stereographic projection 
of the black hole's shadow onto the observer's sky, the boundary of the shadow 
can be described in these celestial coordinates as follows
\begin{equation}\label{alphabetacc}
\begin{split}
X=\lim_{r_{\mathrm{o}}\rightarrow\infty}\left[-r_{\mathrm{o}}^{2}\sin\vartheta_{
\mathrm{o}}\,\frac{\mathrm{d}\varphi}{\mathrm{d}r}\bigg|_{(r_{\mathrm{o}},
\vartheta_{\mathrm{o}})}\right]\,,\qquad 
Y=\lim_{r_{\mathrm{o}}\rightarrow\infty}\left[r_{\mathrm{o}}^{2}\,\frac{\mathrm{
d}\theta}{\mathrm{d}r}\bigg|_{(r_{\mathrm{o}},
\vartheta_{\mathrm{o}})}\right]\,,
\end{split}
\end{equation}
where $(r_{\mathrm{o}}, \vartheta_{\mathrm{o}})$ represent the coordinates of 
the distant observer who sights the photons along the inclination angle 
$\vartheta_{\mathrm{o}}$ coming from the light source deflected by the central 
rotating black hole, as seen in Fig. \ref{Fig5}.
The rotating black hole is a spacetime that is stationary, axisymmetric, and 
asymptotically flat. For an observer located far away from the black hole, 
equations \eqref{alphabetacc} give the following results
\begin{equation}\label{alphabetaasycc}
X=-\frac{\xi_{\mathrm{crit}}}{\sin\vartheta_{\mathrm{o}}}\,,\qquad 
Y=\pm\sqrt{\zeta_{\mathrm{crit}}+a^{2}\cos^{2}\vartheta_{\mathrm{o}}-\xi_{
\mathrm{crit}}^{2}\cot^{2}\vartheta_{\mathrm{o}}}\,.
\end{equation}

\begin{figure}[htb]
\centering
	\includegraphics[width=0.33\textwidth]{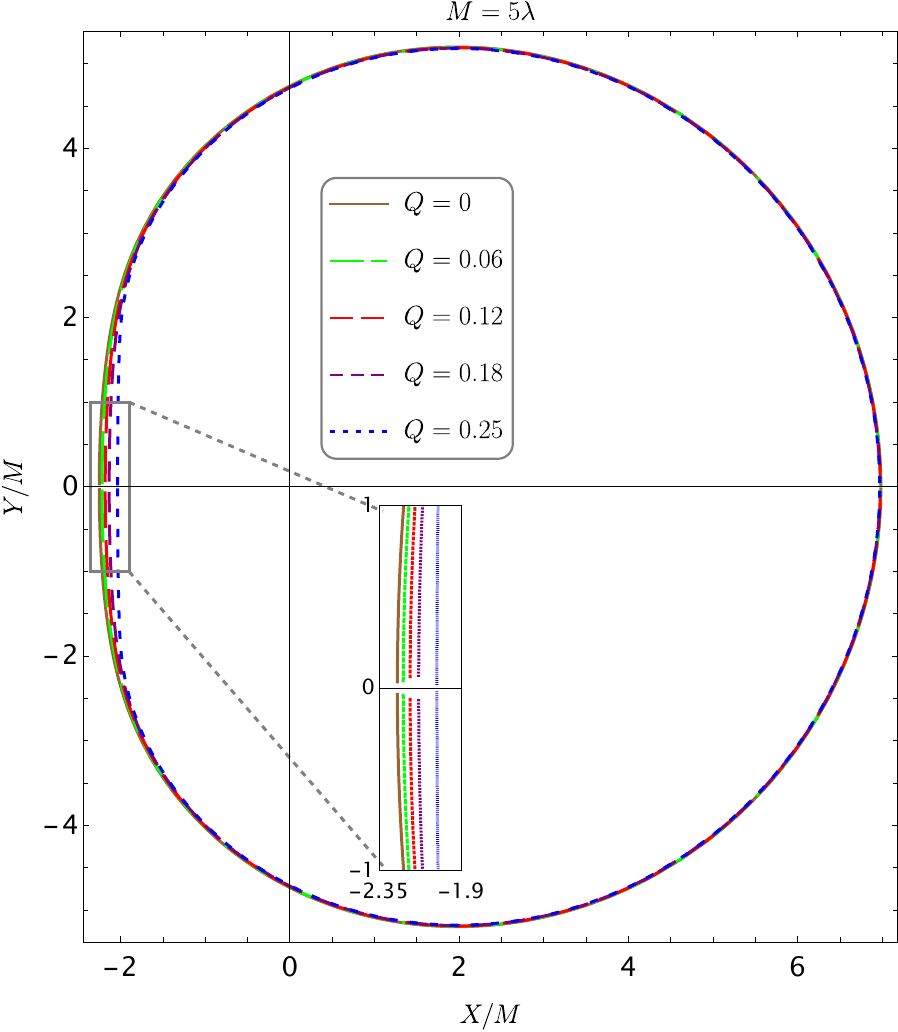}
\caption{\label{Fig8}\small{\emph{Shadow boundary of the rotating scalar-hairy 
black hole configuration in beyond Horndeski gravity, with 
varying parameter $Q>0$, setting $a=0.99$ and 
$\vartheta_{\mathrm{o}}=\frac{\pi}{2}$.}}}
\end{figure}

\begin{figure}[htb]
\centering
	\begin{tabular}{cc}
		\includegraphics[width=0.32\textwidth]{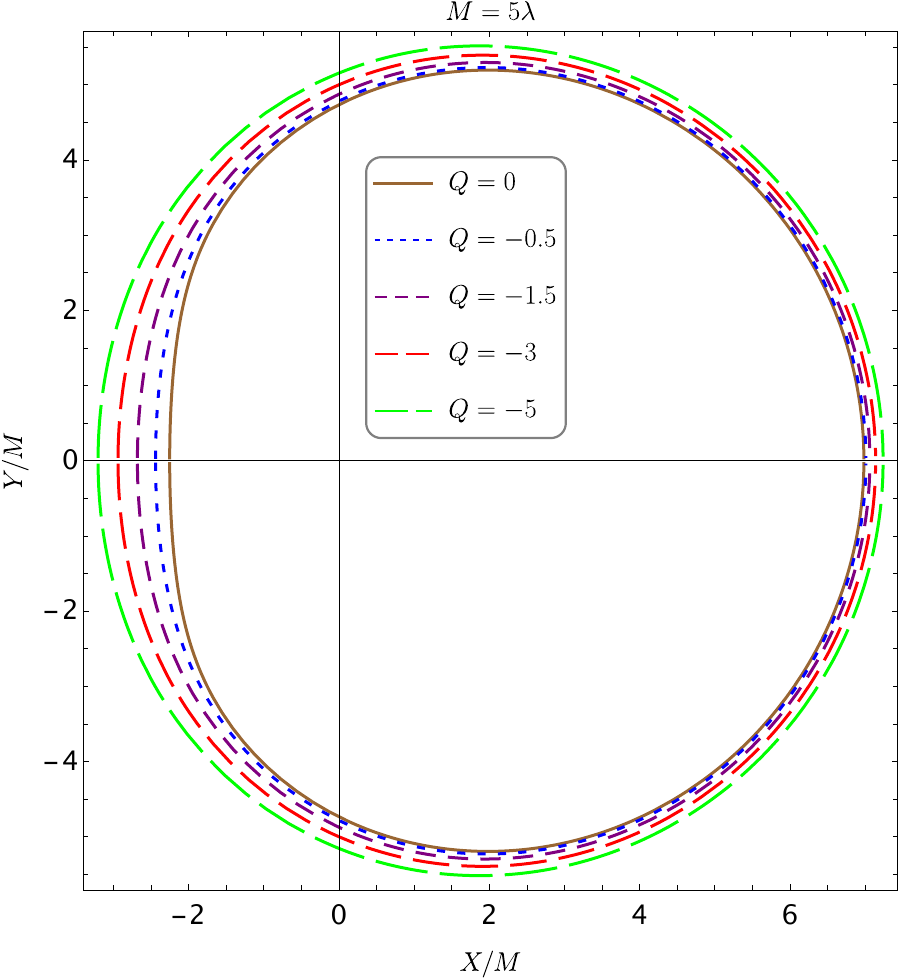}\, &
		\includegraphics[width=0.33\textwidth]{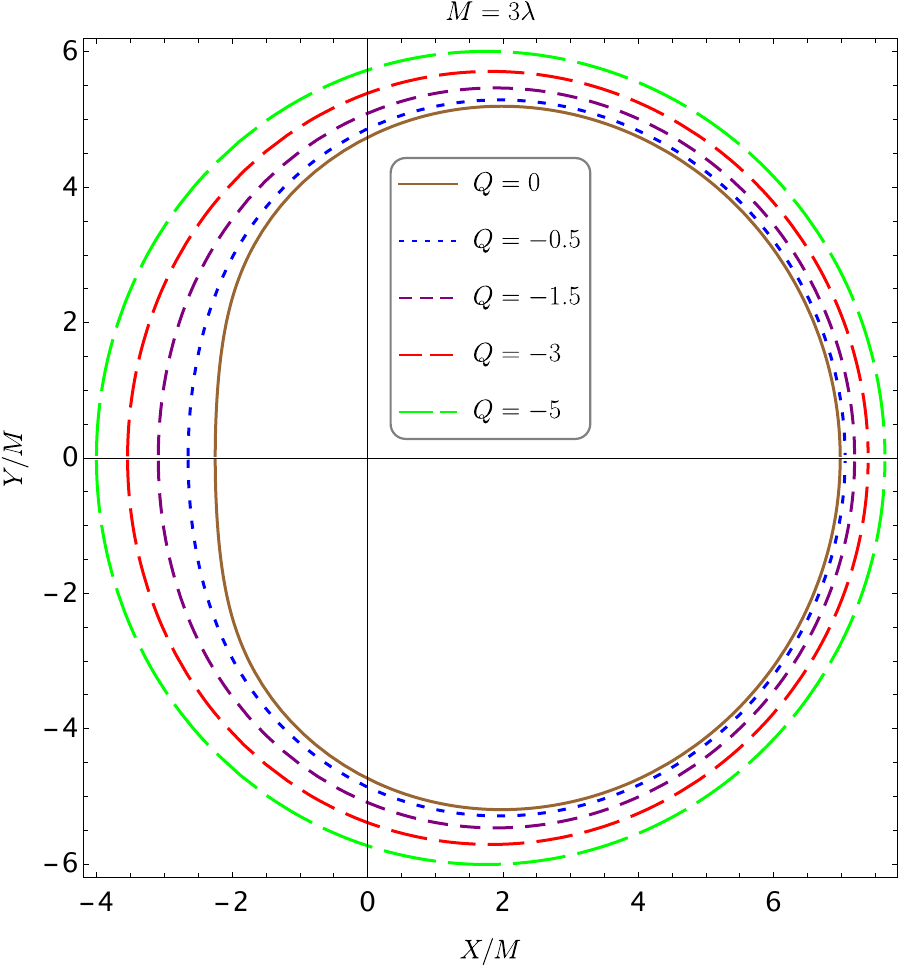}
	\end{tabular}
\caption{\label{Fig9}\small{\emph{Shadow boundary of the rotating scalar-hairy 
black hole configuration in beyond Horndeski gravity, with varying parameter 
$Q<0$, setting $a=0.99$ and $\vartheta_{\mathrm{o}}=\frac{\pi}{2}$.}}}
\end{figure}

\begin{figure}[htb]
\centering
	\begin{tabular}{cc}
		\includegraphics[width=0.37\textwidth]{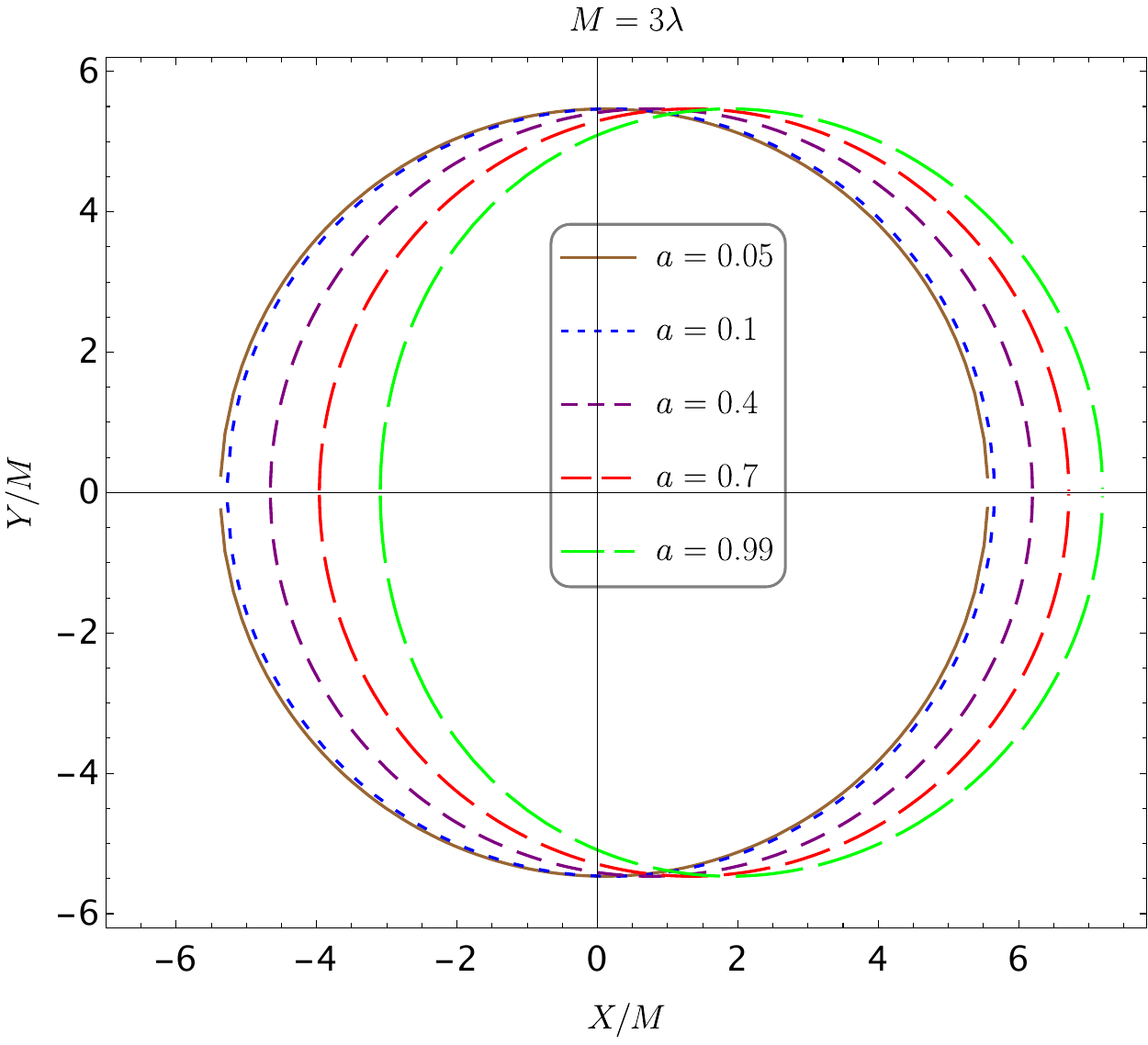}\, &
		\includegraphics[width=0.385\textwidth]{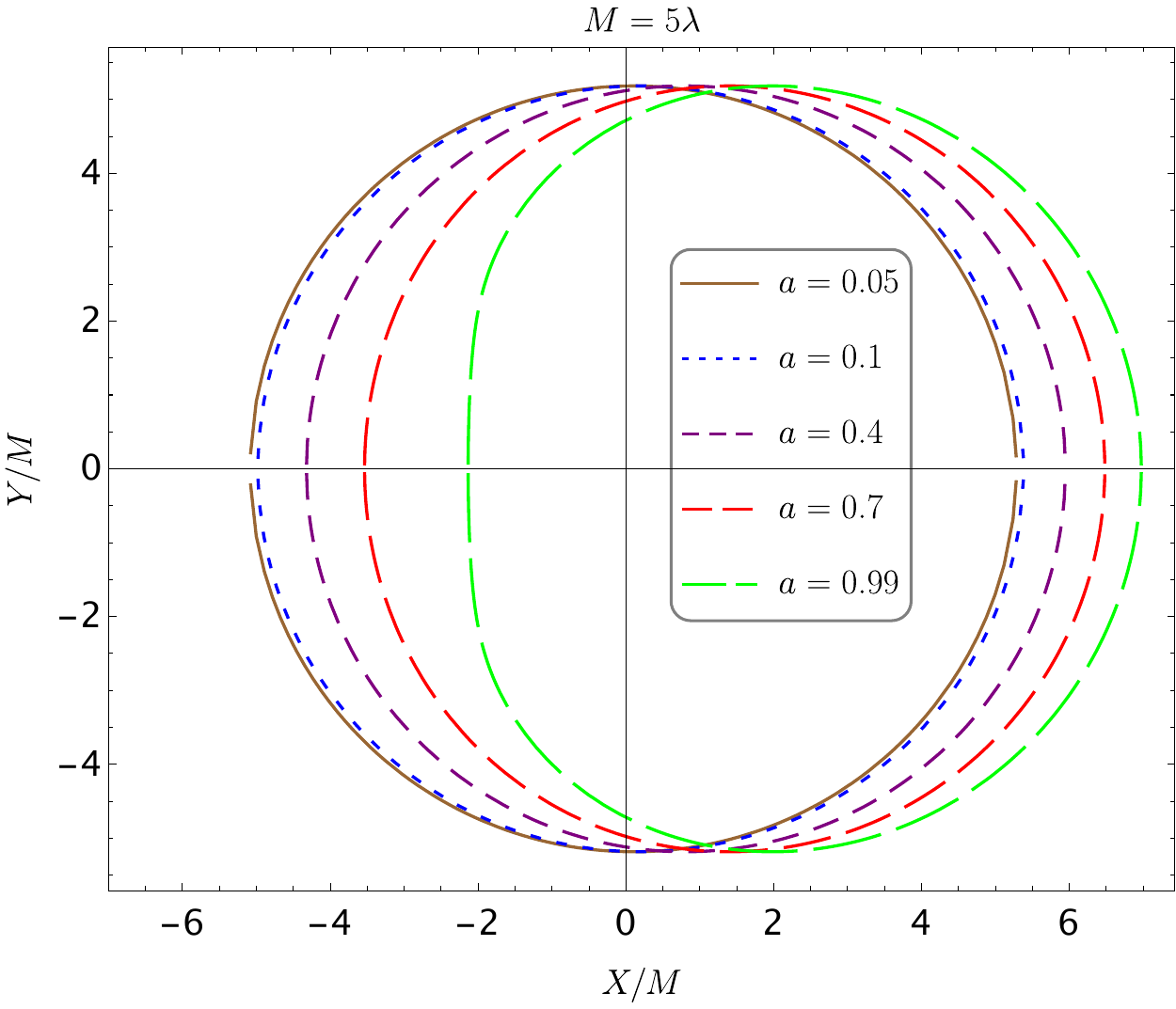}
	\end{tabular}
\caption{\label{Fig10}\small{\emph{Shadow boundary of the rotating scalar-hairy 
black hole configuration in beyond Horndeski gravity, with varying parameter 
$a$, setting $Q=-1.5$ and $\vartheta_{\mathrm{o}}=\frac{\pi}{2}$.}}}
\end{figure}

The $X$ and $Y$ contours define the shadow boundary for rotating black hole 
with scalar hair in beyond Horndeski theory. We present the shadow boundary for 
a range of parameter samples in Figs. \ref{Fig8}, \ref{Fig9}, and \ref{Fig10}. 
As we can see, for $Q<0$ the shadow of a rotating black hole with scalar hair 
in beyond Horndeski theory exhibits a larger area compared to that of the Kerr 
black hole. Conversely, from $Q>0$ the area of the shadow is smaller than that 
of the Kerr black hole. As we observe in Fig. \ref{Fig8}, for $Q>0$, increasing 
the scalar hair parameter $Q$ results in a reduction in the size of the black 
hole's shadow, while simultaneously increasing its oblateness. Conversely, 
from Fig. \ref{Fig9} we deduce that for $Q<0$ a decrease in $Q$ leads to an 
enlargement of the shadow size, accompanied by a reduction in its oblateness. 
Finally, Fig. \ref{Fig10} investigates the impact of the spin parameter $a$, 
demonstrating that for a fixed value of $Q$ an increase in $a$ causes a greater 
distortion of the shadow, with the shadow size diminishing accordingly. These 
observations reveal the intricate interplay between the scalar hair parameter 
$Q$ and the spin parameter $a$ in shaping the rotating black hole with scalar 
hair in beyond Horndeski gravity’s shadow characteristics.

The EHT collaboration, utilizing VLBI technology, captured the event 
horizon-scale image of the supermassive black hole M87* 
\citep{EventHorizonTelescope:2019uob,EventHorizonTelescope:2019dse}. This 
significant observation offers a unique opportunity to probe gravity in the 
strong-field regime and refines our understanding of black holes. The M87* 
shadow appears nearly circular with a crescent shape, exhibiting a circularity 
deviation of $\Delta C\le 0.10$ (10\%), measured as the root-mean-square 
deviation from the average shadow radius. The angular size of the emission 
region, $\theta_d$, in the observed image is $42\pm 3~\mu\mathrm{as}$ 
\citep{EventHorizonTelescope:2019uob,EventHorizonTelescope:2019dse}, and the 
observed image is consistent with the modeled shadow of a Kerr black hole 
\citep{EventHorizonTelescope:2019uob,EventHorizonTelescope:2019dse}, which can 
further be used to place constraints on modified gravity models.

The boundary of a black hole shadow is described by polar coordinates 
$(R(\varphi),\varphi)$, with the origin at the shadow's center $(X_{c},Y_{c})$, 
where $X_{c}=(X_{r}-X_{l})/2$ and $Y_{c}=0$. The shadow reference circle matches 
the shadow contour at the top $(X_{t},Y_{t})$, bottom $(X_{b}, Y_{b})$, and 
rightmost point $(X_{r}, 0)$. The points $(X'_{l}, 0)$ and $(X_{l}, 0)$ denote, 
respectively, the intersections of the reference circle and the leftmost 
boundary of the shadow contour with the horizontal axis. The average shadow 
radius, $\bar{R}$, is defined as \citep{Bambi:2019tjh}
\begin{eqnarray}
\bar{R}^{2}=\frac{1}{2\pi}\int_{0}^{2\pi}R^{2}(\varphi)\mathrm{d}\varphi\,,
\end{eqnarray}
where 
\begin{eqnarray}
R(\varphi)=\sqrt{(X-X_{c})^{2}+(Y-Y_{c})^{2}}\,,\quad \text{and}\quad 
\varphi\equiv\tan^{-1}\left(\frac{Y}{X-X_{c}}\right)\,.
\end{eqnarray}
In this context, $\varphi$ denotes the angle formed between the $x$-axis and the 
vector that links the center $(X_{c},Y_{c})$ to a point $(X,Y)$ located on the 
edge of the shadow. Moreover, the circularity deviation $\Delta C$, i.e., the 
deviation from a perfect circular shape, is quantified by \citep{Bambi:2019tjh} 
\begin{eqnarray}\label{circularity}
\Delta 
C=\frac{1}{\bar{R}}\sqrt{\frac{1}{2\pi}\int_{0}^{2\pi}\left(R(\varphi)-\bar{R}
\right)^{2}\mathrm{d}\varphi}\,.
\end{eqnarray}
This measure will be valuable for comparing theoretical predictions of rotating 
black hole with scalar hair in beyond Horndeski gravity shadows with the 
observations made by EHT \citep{Bambi:2019tjh}. Finally, we characterize the overall size of the shadow through its
area-equivalent radius. For a shadow symmetric with respect to the
horizontal axis, the shadow area can be written as
\begin{equation}
A_{\rm sh}
=
2\int_{r_p^{\min}}^{r_p^{\max}}
Y(r_p)
\left|
\frac{dX(r_p)}{dr_p}
\right|dr_p ,
\end{equation}
where the absolute value ensures a positive area independently of the
orientation used to parametrize the shadow boundary. We then define the
area-equivalent shadow radius as
\begin{equation}
R_{\rm sh}
\equiv
\sqrt{\frac{A_{\rm sh}}{\pi}}
=
\sqrt{
\frac{2}{\pi}
\int_{r_p^{\min}}^{r_p^{\max}}
Y(r_p)
\left|
\frac{dX(r_p)}{dr_p}
\right|dr_p
}.
\end{equation}
Accordingly, the corresponding angular diameter for an observer at
distance $d$ is
\begin{equation}
\theta_d
=
\frac{2R_{\rm sh}}{d}.
\label{angular_diameter}
\end{equation}
Thus, $R_{\rm sh}$ is a radius-like, area-equivalent shadow scale,
whereas $2R_{\rm sh}$ is the corresponding effective shadow diameter.

As a consistency check, in the Schwarzschild limit the shadow is circular
with critical impact parameter $R_{\rm sh}=3\sqrt{3}M$, and the above
definition therefore gives
\begin{equation}
\theta_d^{\rm Schw}
=
\frac{6\sqrt{3}M}{d}.
\end{equation}
For the adopted M87* values
$M=6.5\times10^{9}M_{\odot}$ and $d=16.8~{\rm Mpc}$, this corresponds
to approximately $39.7~\mu{\rm as}$, consistent with the angular-diameter
scale relevant for comparison with the EHT measurement.

By modeling M87* as a rotating black hole with primary scalar hair in beyond 
Horndeski theory, we can compute the circularity deviation $\Delta C$ based on 
the rotating metric given in \eqref{rle}, and then use the results from the EHT 
observations to impose constraints on the model parameters. The EHT observations 
indicated that $\Delta C\leq 0.1$. To more accurately reference the EHT 
observations, we will assume 
$\vartheta_{\mathrm{o}}=\vartheta_{\mathrm{jet}}=17^{\circ}$ for M87*, as the 
jet inclination relative to the line-of-sight for M87* is estimated to be 
$17^\circ$ \citep{CraigWalker:2018vam}.

A reliable estimate of the mass of M87* is essential for translating shadow 
observables into constraints on the underlying spacetime parameters. Although 
independent mass determinations do not yet fully coincide 
\citep{Vagnozzi:2020quf}, we adopt for the present analysis the EHT value 
$M\simeq 6.5\times 10^{9}M_{\odot}$ 
\citep{EventHorizonTelescope:2019uob,EventHorizonTelescope:2019dse}, together 
with a source distance $d=16.8~\mathrm{Mpc}$. 

In addition to the parameters $a$ and $Q$, the dimensionless geometry
depends on the ratio $\lambda/M$. In the numerical analysis presented in
Figs.~11--13, we fix
\begin{equation}
M=3\lambda,
\qquad\text{or equivalently}\qquad
\frac{\lambda}{M}=\frac{1}{3}.
\end{equation}
Thus, the $(a,Q)$ parameter-space plots shown below correspond to a
fixed-$\lambda/M=1/3$ slice of the full parameter space. The parameter
$\lambda/M$ is not varied or marginalized over in the present analysis.

The dependence on $\lambda/M$ follows directly from the appearance of
$r/\lambda$ in the metric function $f(r)$. Introducing
$x=r/M$ and $\ell=\lambda/M$, one has
\begin{equation}
f(x)
=
1-\frac{2}{x}
+
Q\left[
\frac{1}{1+(x/\ell)^2}
+
\frac{\frac{\pi}{2}-\arctan(x/\ell)}
{x/\ell}
\right],
\end{equation}
so that the dimensionless rotating geometry depends on
$(a/M,Q,\ell)$. Consequently, different choices of $\lambda/M$ would in
general modify $\Delta$, the radial potential, the photon region, and the
resulting shadow observables. The present EHT analysis is therefore
conditional on the representative choice $\lambda/M=1/3$.

\begin{figure}[htb]
\centering
	\begin{tabular}{cc}
        \includegraphics[width=0.43\textwidth]{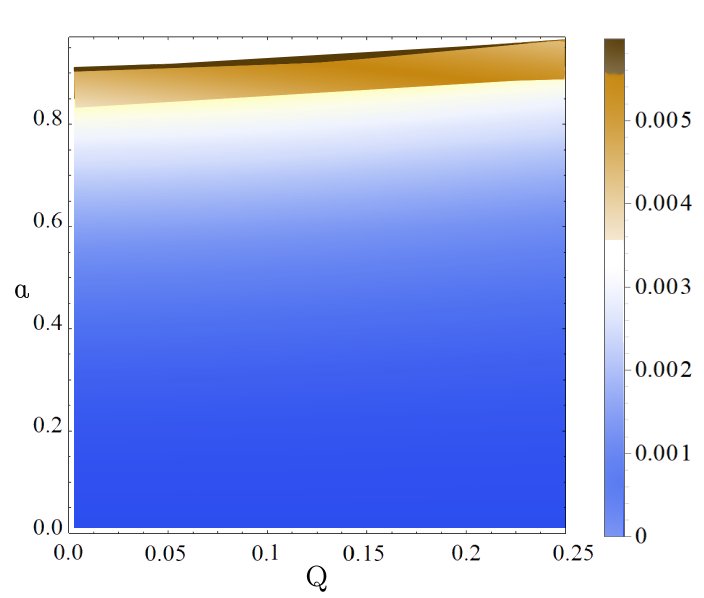}\, &
		\includegraphics[width=0.425\textwidth]{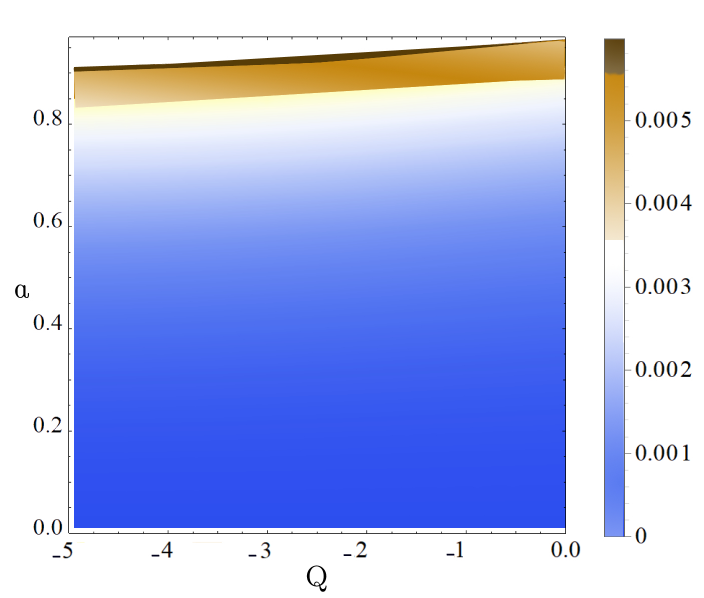}
	\end{tabular}
\caption{\label{Fig99}\small{\emph{The circularity deviation $\Delta C$ for  
rotating scalar-hairy black hole configuration in beyond Horndeski 
theory shadow, as a function of the parameters $a$ and $Q$,   is consistent 
with the EHT observational constraint of the M87* black hole, where $\Delta 
C\leq 0.1$ is satisfied across the entire parameter space of $a$ and $Q$. The 
parameters for M87* used in this analysis are $M=6.5\times 10^{9}M_{\odot}$ and 
$d=16.8~\mathrm{Mpc}$, and the inclination angle considered is 
$\vartheta_{\mathrm{o}}=17^{\circ}$ and $\lambda/M=1/3$. The white region represents the forbidden 
parameter space.}}}
\end{figure}

 In Fig.~\ref{Fig99}, we present 
the circularity deviation $\Delta C$ for an observer inclination angle 
$\vartheta_{\mathrm{o}}=17^{\circ}$. We find that the condition $\Delta C\le 
0.10$ is satisfied throughout the explored $(a,Q)$ parameter space, indicating 
that the current circularity constraint does not exclude rotating black holes 
with primary scalar hair in beyond-Horndeski gravity for this viewing 
configuration.

We now proceed to calculate the angular diameter of the shadow using
Eq.~\eqref{angular_diameter}. Here, $R_{\rm sh}$ denotes the
area-equivalent shadow radius, while the physical diameter entering the
comparison with the EHT measurement is $2R_{\rm sh}$. In Fig.~12, we
show the resulting angular diameter as a function of $a$ and $Q$. The
green curve corresponds to $\theta_d=39~\mu{\rm as}$, namely the lower
edge of the $1\sigma$ EHT interval
$\theta_d=42\pm3~\mu{\rm as}$ for M87*. As throughout the present EHT
analysis, we fix $\lambda/M=1/3$.
\begin{figure}[htb]
\centering
	\begin{tabular}{cc}
        \includegraphics[width=0.43\textwidth]{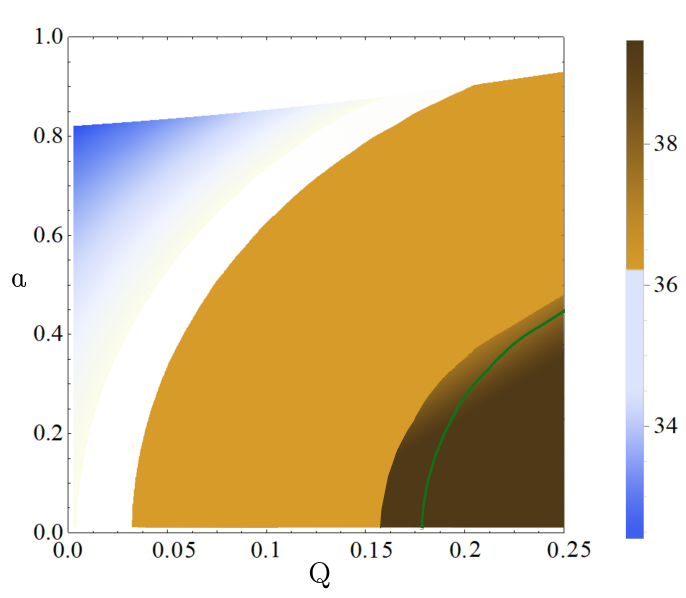}\, &
		\includegraphics[width=0.425\textwidth]{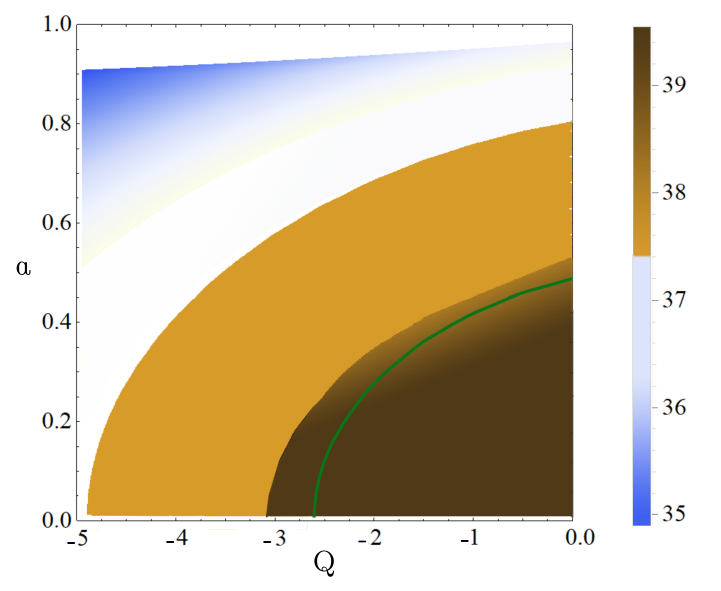}
	\end{tabular}
\caption{\label{Fig100}\small{\emph{The angular diameter
$\theta_d=2R_{\rm sh}/d$ of the phenomenological rotating
scalar-hairy shadow as a function of the parameters $a$ and $Q$, where
$R_{\rm sh}=\sqrt{A_{\rm sh}/\pi}$ is the area-equivalent shadow radius.
The green curve corresponds to $\theta_d=39~\mu{\rm as}$, indicating
the lower bound of the $1\sigma$ observational interval
$\theta_d=42\pm3~\mu{\rm as}$ reported by the EHT. The parameters
adopted for M87* are $M=6.5\times10^{9}M_{\odot}$,
$d=16.8~{\rm Mpc}$, $\vartheta_o=17^\circ$, and
$\lambda/M=1/3$. The white region represents the forbidden parameter
space.}}}
\end{figure}

\begin{figure}[htb]
\centering
	\begin{tabular}{cc}
        \includegraphics[width=0.43\textwidth]{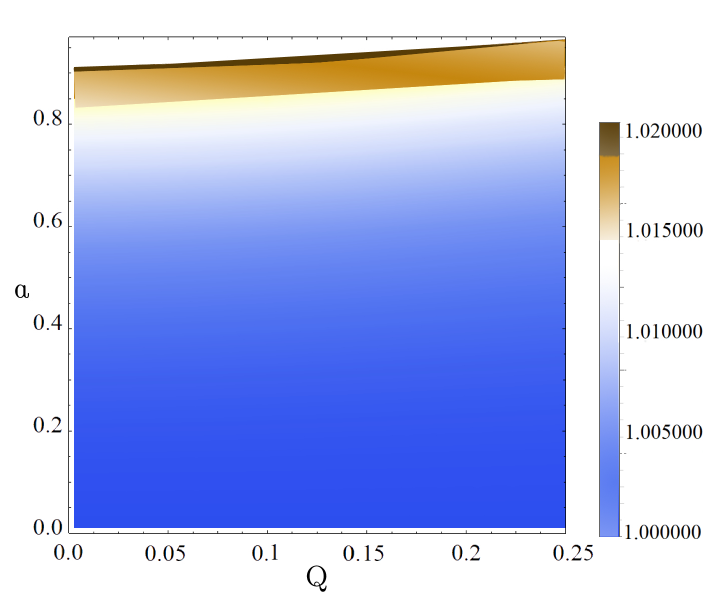}\, &
		\includegraphics[width=0.425\textwidth]{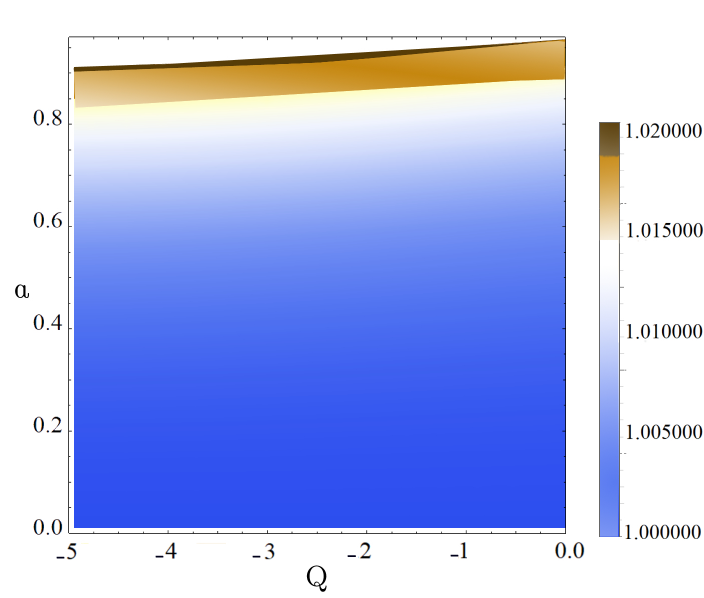}
	\end{tabular}
\caption{\label{Fig101}\small{\emph{The axis ratio observable $D_{x}$ for 
rotating scalar-hairy black hole configuration in beyond Horndeski theory 
shadow as a function of the parameters $a$ and $Q$, at an inclination angle 
$\vartheta_{\mathrm{o}}=17^{\circ}$ and $\lambda/M=1/3$.
Consistently with the EHT observational constraints  for M87*, the condition 
$1<D_{x}\lesssim 4/3$ is satisfied across the entire parameter space. The white 
region represents the forbidden region.}}}
\end{figure}

The circular asymmetry in the M87* shadow can be expressed in terms of the axial 
ratio $D_{x}$, which is defined as the ratio of the major to the minor diameter 
of the shadow 
\citep{EventHorizonTelescope:2019uob,EventHorizonTelescope:2019dse}. The axis 
ratio is given by \citep{Banerjee:2019nnj}
\begin{equation}
D_{x}=\frac{\Delta Y}{\Delta X}\,.
\end{equation}
The axial ratio $D_{x}$ is expected to lie within the range $1<D_{x}\lesssim 
4/3$, in agreement with the EHT observations of M87* 
\citep{EventHorizonTelescope:2019uob,EventHorizonTelescope:2019dse}. In fact, 
$D_{x}$ provides an alternative way to quantify the circularity deviation 
$\Delta C$. The EHT observations reveal that the reconstructed emission ring of 
M87* closely approximates a circle, with an axial ratio of $4{:}3$, which 
corresponds to $\Delta C\le 0.1$ 
\citep{EventHorizonTelescope:2019uob,EventHorizonTelescope:2019dse}. The axial 
ratio is presented as a density plot in Fig.~\ref{Fig101}, where it is clear 
that $1<D_{x}\lesssim 4/3$ holds across the entire parameter space $(Q,a)$. This 
result is in remarkable agreement with the EHT images of M87*, suggesting that 
the observational data from M87* does not rule out rotating black holes with 
scalar hair in beyond Horndeski theory.

\section{Phenomenological implications and parameter sensitivity}\label{sec:4}

Having established the null geodesic structure and the corresponding shadow
construction, we now examine the observational implications of the rotating
primary scalar-hair geometry. The shadow boundary provides a direct mapping
between unstable photon orbits and measurable quantities on the observer's sky,
and therefore constitutes a sensitive probe of deviations from the Kerr
geometry. In the present model, the shadow observables depend on the mass $M$,
spin $a$, effective deformation parameter $Q$, the additional scale ratio $\lambda/M$,
and the observer inclination angle $\vartheta_{\mathrm{o}}$. In particular,
the ratio $\lambda/M$ enters the metric function $f(r)$ and consequently
affects the radial potential, photon region, and predicted shadow size. The
spin primarily introduces asymmetry and displacement, whereas the effective deformation parameter
modifies the effective radial potential of photon motion, thereby altering the
size and distortion of the photon region. In the present EHT analysis, we fix
$\lambda/M=1/3$, equivalently $M=3\lambda$, and investigate the dependence
of the shadow observables on $(a,Q)$ at this fixed value of $\lambda/M$.
Thus, the parameter regions presented below should be understood as
fixed-$\lambda/M$ slices rather than constraints independent of the scale
$\lambda$. This distinction allows, in principle, for observational separation
between rotation-induced and hair-induced effects, although partial
degeneracies may arise.

In what follows, we quantify the sensitivity of the shadow observables to 
variations in $(a,Q)$ and assess their compatibility with current EHT 
measurements. We focus in particular on the angular diameter $\theta_{d}$, the 
circularity deviation $\Delta C$, and the axis ratio $D_{x}$, which provide 
complementary measures of shadow size and distortion and thus offer a direct 
interface between theory and horizon-scale imaging data.

\subsection{Impact of scalar hair on shadow morphology}

The presence of primary scalar hair leads to characteristic modifications of the 
black hole shadow, reflecting the altered structure of null geodesics in the 
strong-field regime. As demonstrated in the previous section, these effects 
manifest in both the overall scale of the shadow and its geometric deformation, 
providing a phenomenological imprint of beyond Horndeski gravity on 
horizon-scale observables.

A salient feature of the shadow morphology is the distinct role played by the 
effective deformation parameter $Q$ compared to the spin parameter $a$. While rotation 
primarily introduces asymmetry and horizontal displacement of the shadow, scalar 
hair predominantly affects its size and degree of circularity. In particular, 
negative-$Q$ deformation systematically increase the shadow radius while 
suppressing shape distortions, whereas positive-$Q$ deformation lead to a reduced 
shadow size accompanied by enhanced deformation. This behavior is evident across 
a wide range of spin parameters and persists for observationally relevant 
inclination angles.

The circularity deviation $\Delta C$, shown in Fig.~\ref{Fig99}, remains below 
the observational bound reported by the EHT throughout the explored parameter 
space. This indicates that the inclusion of primary scalar hair does not 
necessarily produce strongly noncircular shadows, even for moderately large 
deviations from the Kerr geometry. From a phenomenological perspective, this 
result implies that near-circular shadow shapes do not uniquely favor Kerr black 
holes and may also arise in scalar-tensor extensions of general relativity.

In contrast, the angular diameter of the shadow exhibits a stronger sensitivity 
to the scalar hair parameter, as illustrated in Fig.~\ref{Fig100}. Variations in 
$Q$ induce measurable changes in the effective photon sphere radius, leading to 
systematic shifts in the predicted angular size of the shadow. These deviations 
originate from genuine modifications of the radial null effective potential and 
therefore reflect intrinsic geometric departures from Kerr, rather than 
coordinate artifacts. Nevertheless, for the case of M87*, the predicted angular 
diameters remain within current observational uncertainties, underscoring the 
limited resolving power of present data.

Taken together, these results demonstrate that scalar hair leaves a subtle but 
coherent imprint on shadow morphology. The combined behavior of shadow size and 
shape indicates that scalar-tensor corrections primarily rescale the photon 
region while preserving near-circularity for a broad range of parameters. This 
phenomenological pattern distinguishes the effects of scalar hair from those 
induced solely by rotation and motivates the use of multiple shadow observables 
in future tests of gravity.

\subsection{Degeneracies with spin and observational outlook}

A central challenge in interpreting black hole shadow observations is the 
presence of degeneracies among the parameters that govern the spacetime 
geometry. In the rotating scalar-hairy black hole considered here, both the spin 
parameter $a$ and the effective deformation parameter $Q$ influence the shadow size and 
shape, leading to overlapping phenomenological signatures that complicate 
parameter inference.

As illustrated by the distortion parameter and axis-ratio behavior in Fig. 
\ref{Fig101}, different combinations of $a$ and $Q$ can produce shadows with 
comparable degrees of deformation. In particular, an increase in $Q$ within the positive-$Q$ branch can mimic the effect of higher spin by reducing the shadow size and 
enhancing asymmetry, whereas the negative-$Q$ deformation counteracts spin-induced distortions by enlarging and circularizing the shadow. This interplay gives rise 
to a partial degeneracy in shadow observables, whereby distinct regions of the 
$(a,Q)$ parameter space yield nearly indistinguishable shadow morphologies.

This degeneracy is further reflected in the behavior of the circularity 
deviation and angular diameter shown in Figs. \ref{Fig99} and \ref{Fig100}. 
While the circularity deviation remains below the current EHT bound across the 
allowed parameter space, its weak sensitivity limits its ability to disentangle 
scalar hair from rotational effects. The angular diameter provides stronger 
discriminatory power, as it responds more directly to modifications of the 
effective photon sphere radius. However, even this observable is subject to 
degeneracies when uncertainties in the black hole mass, distance, and 
inclination angle are taken into account.

From an observational standpoint, these results indicate that shadow morphology 
alone is insufficient to uniquely constrain the scalar hair parameter at the 
current level of experimental precision. The degeneracy between $a$ and $Q$ 
implies that constraints derived from a single observable may be systematically 
biased if Kerr geometry is assumed \textit{a priori}. A robust test of 
scalar-tensor gravity therefore requires a multi-observable approach that 
combines shadow size, shape, and displacement with independent astrophysical 
information.

Looking ahead, future horizon-scale imaging experiments are expected to 
significantly improve the situation. The next-generation EHT, with enhanced 
baseline coverage and sensitivity, may reduce uncertainties in the shadow 
diameter and shape to a level where subtle deviations induced by scalar hair 
become detectable. In parallel, incorporating complementary probes, such as 
polarization measurements, time-dependent imaging, or constraints from 
accretion-flow dynamics, may help break the degeneracy between spin and scalar 
hair.

In summary, while current EHT observations do not exclude the
phenomenological rotating metric with the primary-hair deformation considered
here, the degeneracies identified in our analysis define clear
phenomenological targets for future observations.

\section{Concluding Remarks}\label{sec:5}

In this work, we have investigated the horizon-scale phenomenology of a
rotating metric constructed from the exact static black-hole solution with
primary scalar hair in beyond Horndeski gravity. Starting from the static
solution, we generated a phenomenological rotating extension via the revised
NJA and analyzed its horizon structure, ergoregion properties, and null
geodesic dynamics. The resulting rotating metric contains the deformation
parameter $Q$, inherited from the primary scalar hair of the exact static
seed solution, in addition to the mass $M$ and spin $a$, and provides a
well-defined metric-level departure from the Kerr geometry. As discussed in
Appendix B, however, establishing this rotating geometry as an exact solution
of the underlying beyond-Horndeski theory would require solving and verifying
the complete metric--scalar field equations.

A central goal of this work was to determine how this primary scalar hair 
modifies the photon region and the observable black hole shadow. By solving the 
Hamilton-Jacobi equations and deriving the critical impact parameters 
analytically, we constructed the full photon region and shadow boundary for 
arbitrary $(a,Q,\vartheta_{\mathrm{o}})$. We showed that scalar hair induces 
systematic and sign-dependent modifications of the shadow morphology. In 
particular, the negative-$Q$ branch ($Q<0$, corresponding to $\eta<0$) enlarges the shadow and reduces its 
oblateness, whereas the positive-$Q$ branch ($Q>0$, corresponding to $\eta>0$) shrinks the shadow and enhances 
its distortion. These effects originate from genuine modifications of the radial 
potential governing null geodesics and therefore reflect intrinsic changes in 
the spacetime geometry rather than coordinate artifacts.

An important result of our analysis is the distinct phenomenological role played 
by scalar hair compared to rotation. While the spin parameter $a$ primarily 
controls asymmetry and displacement of the shadow, the scalar hair parameter $Q$ 
predominantly rescales the photon region and alters the effective photon-sphere 
structure. This difference leads to a characteristic interplay between $(a,Q)$, 
generating partial degeneracies in shadow observables. In particular, certain 
combinations of positive-$Q$ deformation and higher spin can mimic Kerr-like 
distortions, while negative-$Q$ deformation can counteract spin-induced asymmetries. 
A central novelty of the present analysis is the explicit identification and 
quantification of this degeneracy between rotation and primary scalar hair.

Adopting M87* as a representative case within this framework, we 
confronted the model with the current EHT constraints on the circularity 
deviation, angular diameter, and axis ratio. We found that the circularity bound 
$\Delta C\le 0.1$ and the axis ratio constraint $1<D_{x}\lesssim 4/3$ are 
satisfied across the full explored parameter space for 
$\vartheta_{\mathrm{o}}=17^{\circ}$. For the fixed choice $\lambda/M=1/3$ adopted in the M87* analysis, the
angular diameter constraint $\theta_d=42\pm3~\mu{\rm as}$ provides the
strongest restriction on the $(a,Q)$ parameter space, especially for
$Q>0$. The 
scalar-hair-induced deviations in $\theta_{d}$ are typically of order 
$\mathcal{O}(\mu{\rm as})$, placing them close to the current observational 
sensitivity.

These findings show that the phenomenological rotating metric considered here
can remain compatible with the current EHT bounds over a nontrivial region of
its parameter space, while the angular-diameter measurement already provides
meaningful restrictions, particularly for positive values of $Q$. At the same time, 
the small magnitude of the deviations highlights the need for improved angular 
resolution and multi-observable analyses in order to break the degeneracy 
between scalar hair and rotation. Future facilities such as the next-generation 
EHT and space-based VLBI missions are expected to significantly reduce 
uncertainties in shadow diameter and shape, potentially allowing for direct 
discrimination between Kerr and scalar-hairy configurations.

In summary, black-hole shadows provide a powerful probe of deviations from
the Kerr geometry. The phenomenological rotating extension considered in this
work provides a useful metric-level framework for investigating the possible
shadow signatures associated with the primary-hair deformation inherited from
the exact static beyond-Horndeski solution. Establishing whether this rotating
metric can itself be supported by a regular scalar configuration satisfying
the complete beyond-Horndeski field equations remains an important problem for
future work.

\begin{acknowledgments}

The Work of KN is supported financially by the INSF of Iran under the grant number $40501739$, and SS is supported financially by the INSF of Iran under the grant number $40408889$. ENS acknowledges the contribution of the LISA CosWG, and of COST Actions CA18108 ``Quantum Gravity Phenomenology in the multi-messenger approach'' and CA21136 ``Addressing observational tensions in cosmology with systematics and fundamental physics (CosmoVerse)''.

\end{acknowledgments}

\appendix

\section{The revised NJA procedure}\label{apap1}

Following the procedure introduced in Ref. \cite{Azreg-Ainou:2014pra}, we consider the following general static line element
\begin{equation}\label{a1}
\mathrm{d}s^{2}=-G(r)dt^{2}+\frac{\mathrm{d}r^{2}}{F(r)}+H(r)\left(\mathrm{d}\theta^{2}+\sin^{2}\theta\mathrm{d}\phi^{2}\right)\,.
\end{equation}
We apply the advanced null coordinates $(u,r,\theta,\phi)$ defined by
\begin{equation}
\mathrm{d}t=\mathrm{d}r\sqrt{\frac{1}{F(r)G(r)}}+\mathrm{d}u\,,
\end{equation}
to find the following nonzero components of the resulting inverse metric tensor
\begin{equation}
g^{\mu\nu}=-\left(l^{\mu}n^{\nu}+l^{\nu}n^{\mu}-m^{\nu}\bar{m}^{\mu}-m^{\mu}\bar{m}^{\nu}\right)\,,
\end{equation}
where we have introduced
\begin{equation}
g^{\mu\nu}=-\left(l^{\mu}n^{\nu}+l^{\nu}n^{\mu}-m^{\nu}\bar{m}^{\mu}-m^{\mu}\bar{m}^{\nu}\right)\,,
\end{equation}
in which
\begin{equation}
\begin{split}
& l^{\mu}=\delta_{r}^{\mu}\,, \\
& n^{\mu}=\sqrt{\frac{F(r)}{G(r)}}\delta^{\mu}_{u}-\frac{F(r)}{2}\delta^{\mu}_{r}\,, \\
& m^{\mu}=\frac{1}{\sqrt{2H(r)}}\left(\delta_{\theta}^{\mu}+\frac{\mathrm{i}}{\sin\theta}\,\delta_{\phi}^{\mu}\right)\,, \\
& \bar{m}^{\mu}=\frac{1}{\sqrt{2H(r)}}\left(\delta_{\theta}^{\mu}-\frac{\mathrm{i}}{\sin\theta}\,\delta_{\phi}^{\mu}\right)\,,
\end{split}
\end{equation}
and $l_{\mu}l^{\mu}=m_{\mu}m^{\mu}=n_{\mu}n^{\mu}=l_{\mu}m^{\mu}=n_{\mu}m^{\mu}=0$ and $l_{\mu}n^{\mu}=-m_{\mu}{\bar{m}}^{\mu}=1$.

Next, we proceed to apply the following complex transformation
\begin{equation}\label{n1}
r\to r+\mathrm{i}a\cos\theta\,,\qquad u\to u-\mathrm{i}a\cos\theta\,,
\end{equation}
to the metric tensor by which $\delta_{\nu}^{\mu}$ transform as vectors so that
\begin{equation}\label{tr1}
\begin{split}
& \delta_{r}^{\mu}\to\delta_{r}^{\mu}\,, \\
& \delta_{u}^{\mu}\to\delta_{u}^{\mu}\,, \\
& \delta_{\theta}^{\mu}\to\delta_{\theta}^{\mu}+\mathrm{i}a\sin\theta(\delta_{u}^{\mu}-\delta_{r}^{\mu})\,, \\
& \delta_{\phi}^{\mu}\to\delta_{\phi}^{\mu} \,.
\end{split}
\end{equation}
Moreover, we consider that
\begin{equation}\label{tr2}
(G(r),F(r),H(r))\to(A(r,\theta,a),B(r,\theta,a),\Psi(r,\theta,a))\,,
\end{equation}
where $(A(r,\theta,a),B(r,\theta,a),\Psi(r,\theta,a))$ are unknown three-variable real functions. Finally, in order to recover the line element \eqref{a1} in the limit $a\to 0$, we assume that
\begin{equation}\label{n3}
\begin{split}
& \lim_{a\to 0}A(r,\theta,a)=G(r)\,, \\
& \lim_{a\to 0}B(r,\theta,a)=F(r)\,, \\
& \lim_{a\to 0}\Psi(r,\theta,a)=H(r)\,,
\end{split}
\end{equation}

We diverge from the usual NJA approach, which determines the expressions for $(A(r,\theta,a),B(r,\theta,a),\Psi(r,\theta,a))$ by complexifying the radial coordinate $r$. Instead, in our method, we will establish $(A(r,\theta,a),B(r,\theta,a),\Psi(r,\theta,a))$ based on different criteria and physical arguments. Therefore, the complex transformation \eqref{n1} will lead to the following expressions
\begin{equation}
\begin{split}
& l^{\mu}=\delta_{r}^{\mu}\,, \\
& n^{\mu}=\delta_{u}^{\mu}\sqrt{\frac{B(r,\theta,a)}{A(r,\theta,a)}}-\frac{1}{2}\delta_{r}^{\mu}B(r,\theta,a)\,, \\
& m^{\mu}=\frac{1}{\sqrt{2\Psi(r,\theta,a)}}\left(\delta_{\theta}^{\mu}+\mathrm{i}a\sin\theta\left(\delta_{u}^{\mu}-\delta_{r}^{\mu}\right)+\frac{\mathrm{i}
\delta_{\phi}^{\mu}}{\sin\theta}\right)\,, \\
& \bar{m}^{\mu}=\frac{1}{\sqrt{2\Psi(r,\theta,a)}}\left(\delta_{\theta}^{\mu}-\mathrm{i}a\sin\theta\left(\delta_u^{\mu}-\delta_r^{\mu}\right)-\frac{\mathrm{i}
\delta_{\phi}^{\mu}}{\sin\theta}\right)\,.
\end{split}
\end{equation}
Hence, we can find the non-zero components of the transformed inverse metric tensor as follows
\begin{equation}
\begin{split}
& g^{uu}=\frac{a^{2}\sin^{2}(\theta)}{\Psi(r,\theta,a)}\,,\qquad g^{u\phi}=\frac{a}{\Psi(r,\theta,a)}\,,\qquad g^{r\phi}=-\frac{a}{\Psi(r,\theta,a)}\,,\\
& g^{rr}=\frac{a^{2}\sin^{2}(\theta)}{\Psi(r,\theta,a)}+B(r,\theta,a)\,,\\
& g^{\theta\theta}=\frac{1}{\Psi(r,\theta,a)}\,,\qquad g^{\phi\phi}=\frac{1}{\sin^{2}(\theta)\Psi(r,\theta,a)}\,,\\
& g^{ur}=-\left(\frac{a^{2}\sin^{2}\theta}{\Psi(r,\theta,a)}+\sqrt{\frac{B(r,\theta,a)}{A(r,\theta,a)}}\right)\,,
\end{split}
\end{equation}
and consequently, the rotating line element in EFCs takes the form
\begin{equation}\label{eeee}
\begin{split}
\mathrm{d}s^{2} & =\mathrm{d}u^{2}(-A(r,\theta,a))-2\frac{\sqrt{A(r,\theta,a)}}{\sqrt{B(r,\theta,a)}}\mathrm{d}r\mathrm{d}u-2a\sin^{2}\theta\left(\frac{\sqrt{A(r,\theta,a)}}
{\sqrt{B(r,\theta,a)}}-A(r,\theta,a)\right)\mathrm{d}u\mathrm{d}\phi \\
& +\frac{2a\sin^{2}\theta\sqrt{A(r,\theta,a)}}{\sqrt{B(r,\theta,a)}}\mathrm{d}r\mathrm{d}\phi+\Psi(r,\theta,a)\mathrm{d}\theta^{2} \\
& +\sin^{2}\theta\left(a^{2}\sin^{2}\theta\left(\frac{2\sqrt{A(r,\theta,a)}}{\sqrt{B(r,\theta,a)}}-A(r,\theta,a)\right)+\Psi(r,\theta,a)\right)\mathrm{d}\phi^{2}\,.
\end{split}
\end{equation}

The crucial step is to transform the line element \eqref{eeee} into BLCs through a global coordinate transformation, typically represented in the following form
\begin{equation}\label{jjjj}
\mathrm{d}u=\mathrm{d}t+\lambda (r)\mathrm{d}r \,,\qquad 
\mathrm{d}\phi=\mathrm{d}\varphi+\chi (r)\mathrm{d}r\,.
\end{equation}
In order to ensure the integrability of Eq. \eqref{jjjj}, the functions $\lambda(r)$ and $\chi(r)$ must depend solely on $r$.

The usual NJA typically fails to transform the line element \eqref{eeee} into BLCs. This is because, in the NJA, $(A(r,\theta,a),B(r,\theta,a),\Psi(r,\theta,a))$ are determined by the complexification of $r$, leaving no free functions to facilitate the transformation into BLCs. However, this limitation does not apply to the revised NJA method, as the functions $(A(r,\theta,a),B(r,\theta,a),\Psi(r,\theta,a))$ remain unknown. Thus, we can successfully achieve the transformation into BLCs. Taking this into account, we proceed with
\begin{equation}\label{zzzz}
\lambda(r)=-\frac{K(r)+a^{2}}{a^{2}+F(r)H(r)}\,,\qquad \chi(r)=-\frac{a}{a^{2}+F(r)H(r)}\,,
\end{equation}
where
\begin{equation}\label{zzzz}
K(r)=H(r)\sqrt{\frac{F(r)}{G(r)}}\,.
\end{equation}
Thus, one can obtain
\begin{equation}\label{zzzz}
\begin{split}
& A(r,\theta,a)=\frac{\Psi(r,\theta,a)\left(a^{2}\cos^{2}\theta+F(r)H(r)\right)}{\left(K(r)+a^{2}\cos^{2}\theta\right)^{2}}\,,\\
& B(r,\theta,a)=\frac{a^{2}\cos^{2}\theta+F(r)H(r)}{\Psi(r,\theta,a)}\,,
\end{split}
\end{equation}
leading to the line element \eqref{rle}.

\section{Consistency conditions for the Newman--Janis-generated rotating
geometry}\label{apapc2}

In this Appendix, we examine the consistency of the rotating black hole line 
element \eqref{rle} obtained via the revised NJA with the field equations of the 
beyond Horndeski theory to assess whether the revised NJA can be consistently 
applied in this framework. We show that the scalar field ansatz \eqref{scfi} 
leads to an inconsistency. Hence, we formulate the general conditions required 
for the rotating line element \eqref{rle} to be consistently supported by 
the field equations. These conditions reduce to a coupled system of 
two-dimensional partial differential equations for the scalar field whose 
numerical solution, though nontrivial, would complete the construction of a 
rotating black hole with primary scalar hair, described in the line element 
\eqref{rle}, within this theory. However, the numerical integration of this 
coupled system is complex and beyond the scope of this study.

\subsection{Field equations of the beyond Horndeski gravity}

We proceed by obtaining the field equations of the beyond Horndeski theory, 
restricting attention to the vacuum sector. The field equations are derived by 
independently varying the beyond Horndeski action 
$\mathcal{I}\left[\tilde{g}_{\mu\nu},\Phi\right]$ introduced in 
Eq.~\eqref{action} with respect to the metric tensor $\tilde{g}_{\mu\nu}$ and 
the scalar field $\Phi$, treating them as independent dynamical variables of the 
setup. The metric variation produces the gravitational field equations (modified 
Einstein equations), whereas the scalar variation yields an independent equation 
governing the dynamics of the scalar field $\Phi$. Therefore, the gravitational 
field equations can be found as follows
\begin{eqnarray}\label{gfebht}
\mathcal E_{\mu\nu}&\equiv&
G_4(W)\,\tilde G_{\mu\nu}+\frac{1}{2}G_{2}(W)\,\tilde 
g_{\mu\nu}-(\partial_{W}G_{2}(W))\,\partial_{\mu}\Phi\,\partial_{\nu}
\Phi+2(\partial_{W}G_{4}(W))\Bigl[
(\nabla_{\mu}\nabla_{\nu}\Phi)\,\Box\Phi-\nabla_{\mu}\nabla_{\alpha}\Phi\,
\nabla_{\nu}\nabla^{\alpha}\Phi\Bigr]\nonumber\\
&&
-\,\tilde 
g_{\mu\nu}\,(\partial_{W}G_{4}(W))\Bigl[(\Box\Phi)^{2}-\nabla_{\alpha}\nabla_{
\beta}\Phi\,\nabla^{\alpha}\nabla^{\beta}\Phi\Bigr]+2\,\nabla_{\alpha}
\nabla_{\beta}\Bigl(F_4(W)\,\epsilon^{\alpha\rho\sigma}{}_{(\mu}\,\epsilon^{
\beta\gamma\delta}{}_{\nu)}\,\partial_{\rho}\Phi\,\partial_{\gamma}\Phi\,
\nabla_{\sigma}\nabla_{\delta}\Phi\Bigr)=0\,,
\end{eqnarray}
where the Einstein tensor is $\tilde G_{\mu\nu}=\tilde 
R_{\mu\nu}-\frac{1}{2}\tilde R\tilde g_{\mu\nu}$.

Due to the shift-symmetry of $\Phi$ in the setup, there exists a Noether current
\begin{eqnarray}\label{ncj}
J^{\mu}=\frac{1}{\sqrt{-\tilde{g}}}\frac{\delta\mathcal{I}\left[\tilde{g}_{
\mu\nu},\Phi\right]}{\delta(\partial_{\mu}\Phi)}
&=&
-\,(\partial_{W}G_{2}(W))\,\partial^{\mu}\Phi+\,2\,(\partial_{W}G_{4}(W))\Bigl[
(\Box\Phi)\,\partial^{\mu}\Phi-\nabla^{\mu}\nabla^{\nu}\Phi\,\partial_{\nu}
\Phi\Bigr]
\nonumber\\
&&
-\,2\,F_4(W)\,\epsilon^{\mu\nu\rho\sigma}\epsilon^{\alpha\beta\gamma}{}_{\sigma}
\,\partial_{\nu}\Phi\,\nabla_{\alpha}\nabla_{\beta}\Phi\,\nabla_{\rho}\nabla_{
\gamma}\Phi\,.
\end{eqnarray}
Hence, the scalar field equation can be found
\begin{equation}\label{sfencj}
\nabla_{\mu}J^{\mu}=0\,.
\end{equation}
Eqs. \eqref{gfebht} and \eqref{sfencj} are the full field equations of the 
beyond Horndeski gravity.

One can show that with the static and spherically symmetric black hole 
possessing primary scalar hair \eqref{ssle} and the scalar field ansatz 
\eqref{scfi}, the only independent field equation is the Noether current 
component $J^{r}(r)=0$, which can be fully satisfied. Moreover, the 
associated kinetic term of the scalar field as obtained in Eq. \eqref{Xbh1} is 
merely a function of $r$. Thanks to these conditions, the spherically symmetric 
black hole with primary scalar hair \eqref{ssle} constitutes an exact 
solution of the field equations \eqref{gfebht} and \eqref{sfencj} of the 
beyond Horndeski setup.

\subsection{Generalized scalar ansatz and consistency conditions}

Although the rotating black hole with primary scalar hair \eqref{rle} is 
obtained via the revised NJA, it does not satisfy the field equations 
\eqref{gfebht} and \eqref{sfencj} of the beyond Horndeski theory. If we again 
consider the scalar field ansatz \eqref{scfi}, then the associated kinetic term 
of the scalar field $W=-\frac{1}{2}\partial_{\mu}\Phi\,\partial^{\mu}\Phi$ is no 
longer solely a function of $r$. Consequently, two components of gravitational 
field equations of the beyond Horndeski theory, i.e., $\mathcal E_{r\theta}$ and 
$\mathcal E_{t\theta}$ no longer vanish, so that $\mathcal E_{\mu\nu}\neq 0$. On 
the other hand, we see that $J^{r}(r)\to J^{r}(r,\theta)$. Hence, one can deduce 
that $\nabla_{\mu}J^{\mu}\neq 0$. Therefore, the rotating black hole with 
primary scalar hair \eqref{rle} obtained by the revised NJA is no longer an 
exact solution of the field equations \eqref{gfebht} and \eqref{sfencj} of the 
beyond Horndeski theory. Therefore, the revised NJA does not map static hairy 
black holes into rotating solutions in this theory by considering the scalar 
field ansatz \eqref{scfi}. 

We want to formulate a new ansatz for the scalar field such that the rotating 
black hole with primary scalar hair \eqref{rle} deduced via the revised NJA 
can be consistently supported by the field equations of the beyond 
Horndeski theory. This new ansatz for the scalar field can be defined as follows
\begin{equation}\label{newsca}
\tilde{\Phi}(t,r,\theta)=qt+\tilde{\Psi}(r,\theta)\,.
\end{equation}
The new ansatz for the scalar field \eqref{newsca} means that the scalar field 
carries a conserved, time-linear scalar charge (set by $q$), while its spatial 
profile dynamically readjusts in a stationary and axisymmetric way to 
consistently account for the effects of rotation. Geometrically, the scalar 
defines preferred foliation surfaces whose time direction remains aligned with 
the stationary Killing vector, while the level surfaces of the scalar are 
deformed away from spherical symmetry into axisymmetric ones, reflecting the 
underlying rotating geometry. The associated kinetic term of the scalar field is 
$\tilde{W}(r,\theta)=-\frac{1}{2}g^{\mu\nu}\partial_{\mu}\tilde{\Phi}\partial_{
\nu}\tilde{\Phi}$. Using the inverse metric components
\begin{equation}
g^{tt}
=
-\frac{(r^{2}+a^{2})^{2}-a^{2}\Delta\sin^{2}\theta}
{\Sigma\Delta},
\qquad
g^{rr}=\frac{\Delta}{\Sigma},
\qquad
g^{\theta\theta}=\frac{1}{\Sigma},
\end{equation}
and noting that $\partial_{\phi}\widetilde{\Phi}=0$, the definition
\ref{wtilde-main} gives
\begin{equation}\label{PDE-main2}
\Delta(\partial_{r}\widetilde{\Psi})^{2}
+(\partial_{\theta}\widetilde{\Psi})^{2}
-\frac{(r^{2}+a^{2})^{2}
-a^{2}\Delta\sin^{2}\theta}{\Delta}\,q^{2}
+2\widetilde{W}\Sigma=0.
\end{equation}

As a check, in the static limit $a\to0$, for which
$\Sigma\to r^{2}$ and $\Delta\to r^{2}f(r)$, Eq.~\eqref{PDE-main2}
reduces to
\begin{equation}
f(r)\,[\Psi'(r)]^{2}
-\frac{q^{2}}{f(r)}
+2W(r)=0,
\end{equation}
which is consistent with the kinetic term of the exact static scalar
configuration given in Eqs.~\eqref{scfi}--\eqref{Xbh1}.
The next two equations arise from gravitational field equations as
\begin{equation}\label{ertheta}
\mathcal 
E_{r\theta}=\eta\sin^{2}\theta\left(\Delta(\partial_{r}\partial_{\theta}\tilde{
\Psi})-\Sigma(\partial_{r}\tilde{W})(\partial_{\theta}\tilde{W})\right)=0\,,
\end{equation}
and
\begin{equation}\label{errethetatheta}
\mathcal E_{rr}-\mathcal 
E_{\theta\theta}=\eta\left(\Delta(\partial_{r}^{2}\tilde{W})+(\partial_{\theta}^
{2}\tilde{W})+\frac{2r\Delta}{\Sigma}(\partial_{r}\tilde{W})
-\frac{2a^{2}\sin\theta\cos\theta}{\Sigma}(\partial_{\theta}\tilde{W})-\frac{
\Delta}{\Sigma^{2}}\left[\Delta(\partial_{r}\tilde{\Psi})^{2}-(\partial_{\theta}
\tilde{\Psi})^{2}\right]\right)=0\,.
\end{equation}
Equations~\eqref{PDE-main2}--\eqref{errethetatheta} provide examples of the
nontrivial conditions that arise when the scalar sector is generalized from
the static ansatz to a stationary and axisymmetric configuration. The first
relation follows directly from the definition of the scalar kinetic term,
whereas the latter two arise from selected components and combinations of the
gravitational field equations.

We emphasize that these equations are not claimed here to constitute a
complete or independent set of equations sufficient to establish the
existence of a rotating solution. In particular, we have not demonstrated
that all remaining components of $\mathcal E_{\mu\nu}=0$ follow from the
displayed relations, nor that the scalar-current equation
$\nabla_{\mu}J^{\mu}=0$ is automatically implied by them. Establishing such
relations would require an explicit analysis of the complete system of
metric and scalar equations for the generalized scalar configuration.

The purpose of the above equations is therefore more limited: they
demonstrate explicitly that retaining the static scalar profile after the
Newman--Janis transformation is insufficient and identify some of the
nontrivial conditions that a generalized stationary and axisymmetric scalar
configuration would have to satisfy.

A complete construction would additionally require appropriate boundary and
regularity conditions. In the static limit, the exact scalar configuration
must be recovered,
\begin{equation}
\lim_{a\to0}\widetilde{\Psi}(r,\theta)
=\Psi_{\rm static}(r),
\qquad
\lim_{a\to0}\widetilde W(r,\theta)
=
\frac{q^{2}/2}{1+(r/\lambda)^{2}}.
\end{equation}
Regularity on the symmetry axis requires
\begin{equation}
\left.\partial_{\theta}\widetilde{\Psi}\right|_{\theta=0,\pi}=0,
\qquad
\left.\partial_{\theta}\widetilde W\right|_{\theta=0,\pi}=0,
\end{equation}
while, if equatorial reflection symmetry is imposed,
\begin{equation}
\left.\partial_{\theta}\widetilde{\Psi}\right|_{\theta=\pi/2}=0,
\qquad
\left.\partial_{\theta}\widetilde W\right|_{\theta=\pi/2}=0.
\end{equation}
At spatial infinity, the scalar configuration should approach the
asymptotic behavior compatible with the exact static solution. Regularity at
the future event horizon must be assessed in coordinates that are regular
across the horizon, rather than by requiring the Boyer--Lindquist radial
derivative of the scalar to remain finite. In particular, scalar invariants
such as
$\widetilde W=-\frac12\nabla_{\mu}\widetilde{\Phi}
\nabla^{\mu}\widetilde{\Phi}$
must remain regular there.

Solving the complete boundary-value problem described above is beyond the
scope of the present work. Accordingly, we do not claim that the metric
\eqref{rle} has been established as an exact rotating black-hole solution of
the underlying beyond Horndeski theory. In the main text, \eqref{rle} is
instead employed as a Newman--Janis-generated phenomenological rotating
extension of the exact static primary-hair geometry. The horizon,
null-geodesic, and shadow properties derived from this metric are therefore
properties of this rotating metric parametrization. Determining whether the
same geometry can be supported by a regular stationary and axisymmetric
scalar configuration satisfying the complete beyond-Horndeski field
equations remains a separate problem that requires further investigation.

 \bibliographystyle{apsrev4-2}
\bibliography{Reff}

\end{document}